\documentclass[a4paper,11pt]{article}
\usepackage[T1]{fontenc}
\usepackage[utf8x]{inputenc}
\usepackage{color}
\usepackage{bm}
\usepackage{stmaryrd, mathrsfs, dsfont, amsmath, latexsym, amsfonts}
\usepackage{youngtab}
\usepackage{hyperref}
\usepackage{verbatim}
\usepackage{cite}
\numberwithin{equation}{section}

\def \de{\partial}
\def \R{{\cal R}}
\def \H{{\cal H}}
\def \J{{\cal J}}

\def \N{{\cal N}}

\def \X{{\cal X}}
\def \B{{\cal B}}
\def \G{{\cal G}}
\def \E{{\cal E}}

\newcommand{\ket}[1]{\left\vert #1 \right\rangle}
\newcommand{\bra}[1]{\left\langle #1 \right\vert}

\begin{document}


\setcounter{page}{1}
\begin{flushright}
LMU-ASC 14/20\\
\end{flushright}
\begin{center}

\vskip 1cm


{\Large \bf  A Worldline Theory for Supergravity}


\vspace{12pt}

Roberto Bonezzi\,$^{1}$, Adiel Meyer\,$^2$ and Ivo Sachs\,$^3$\\

\vskip 25pt

{\em $^1$ \hskip -.1truecm Institute for Physics, Humboldt University Berlin,\\
 Zum Gro\ss en Windkanal 6, D-12489 Berlin, Germany\vskip 5pt }

\vskip 5pt
 {\em $^2$ \hskip -.1truecm 
 Department of Natural Sciences, The Open University of Israel, \\ PO Box 808, Ra’anana 43537, Israel}

\vskip 5pt
 {\em $^3$ \hskip -.1truecm Arnold Sommerfeld Center for Theoretical Physics,
Ludwig-Maximilian-Universit\"at ,\\
Theresienstr. 37, D-80333 M\"unchen, Germany}


\vskip 15pt


{emails: \small{{\tt roberto.bonezzi@physik.hu-berlin.de},  
{\tt Adielmb@gmail.com}, 
{\tt Ivo.Sachs@physik.uni-muenchen.de}}}
\end{center}

\vskip 15pt

\begin{center}
{\large{\it In honor of Samson Shatashvili's 60th birthday}}
\end{center}
\vskip.5cm
\begin{abstract}
    The $\N=4$ supersymmetric spinning particle admits several consistent quantizations, related to the gauging of different subgroups of the $SO(4)$ $R$-symmetry on the worldline. We construct the background independent BRST quantization for all of these choices which are shown to reproduce either the massless NS-NS spectrum of the string, or  Einstein theory with or without the antisymmetric tensor field and/or dilaton corresponding to different restrictions. Quantum consistency of the worldline implies equations of motion for the background which, in addition to the admissible string backgrounds, admit Einstein manifolds with or whithout a  cosmological constant. The vertex operators for the Kalb-Ramond, graviton and dilaton fields are obtained from the linear variations of the BRST charge. They produce the physical states by action on the diffeomorphism ghost states.
\end{abstract}
\newpage
\tableofcontents

\vskip 40pt
\section{Introduction and summary of results}
The fact that critical string theory contains a massless graviton in its spectrum and that the consistency of the worldsheet conformal field theory implies the vacuum Einstein equations are generally considered important consistency tests for string theory to be a theory of quantum gravity. On the other hand, in string theory the graviton always comes together with the dilaton and, depending on the model, also an anti-symmetric Kalb-Ramond tensor field. Another important feature in string theory is the operator state correspondence which asserts that any scattering state can be represented by insertion of a suitable vertex operator on the worldsheet. Finally, the absence of conformal anomalies implies coupled equations for all of these fields. These equations are rather restrictive. In particular, they do not seem to allow a spacetime of positive constant curvature as a background. 

A natural question is then to what extent this structure is unique to string theory and whether some simpler model shares some of these features as well. One such example is the chiral string \cite{Hohm:2013jaa} or the ambitwistor string \cite{Mason:2013sva} which are worldsheet theories that, unlike string theory, do not have any massive states. An even more drastic simplification is to replace the worldsheet by a spinning particle on the worldline \cite{Barducci:1976xq,Brink:1976uf,Dai:2008bh}. A well known fact is that ${\cal N}=4$ worldline supersymmetry is required in order to have a spin $2$ particle in the spectrum. One might think that with such a drastic simplification anything should be possible. Yet, the problem of coupling this theory to gravity consistently has been solved only recently \cite{Bonezzi:2018box}. On the other hand, due to the $SO(4)$ $R$-invariance of the worldline action there is not a unique such theory since it can be projected by gauging various subgroups of $SO(4)$.  This freedom in the choice of projection is an extension of what is known as level matching and un-oriented worldsheets in string theory. The theory described in \cite{Bonezzi:2018box} is maximally projected, which corresponds to gauging all of  $SO(4)$. In that case the graviton is the only remaining degree of freedom in the spectrum. As shown in \cite{Bonezzi:2018box} it can be consistently coupled to gravity provided the background is an Einstein space. Furthermore, the physical gravitons on such a background are obtained by acting with the linearized  BRST charge on a diffeomorphism ghost state. This is the worldline manifestation of the operator state correspondence in string theory. Finally, the correct 3-graviton amplitudes are reproduced in the worldline theory. Thus, gauge invariance of the point particle worldline theory, manifested by the nilpotency of the  BRST operator, implies the equations of motion of Einstein Gravity. 

In the present paper we investigate the quantization of the spinning particle with weaker constraints. We find that the minimal projection, that allows for an interacting theory, is obtained by gauging a $U(1)\times U(1)$ subgroup of $SO(4)$. This is closely related to level truncation and level matching in string theory. So one would expect that this worldline theory is the one that is most closely related to string theory. Indeed we recover the complete massless BV-spectrum of NS-NS sector of the closed superstring in the worldline theory, together with the interaction between the different fields. Perhaps surprisingly this also includes the dilaton which, since it couples to the worldsheet curvature, is usually considered to be a specific feature of string theory. Similarly the Kalb-Ramond field, being a two form, naturally couples to the worldsheet rather than a worldline. However, as explained below, the two complex worldline fermions of the supersymmetric spinning particle already provide such a coupling to the Kalb-Ramond field and furthermore, the nilpotency of the BRST charge which is tantamount to the equations of motion of a general BV action, implies the full non-linear equations of motion. In particular, these constraints are not derived from the absence of the conformal anomaly which plays no role here. 

We then proceed to show that linear variations of the BRST charge around a classical background then yield the unintegrated vertex operators for the graviton and the Kalb-Ramond field as well as the dilaton along the same lines as in \cite{Bonezzi:2018box}, as an extension of \cite{Dai:2008bh}. Finally, the three point amplitudes for various fields are correctly reproduced on the worldline. 

We would also like to comment on the relation of our results to other approaches. In \cite{Adamo:2014wea,Adamo:2017nia} the coupling of the worldsheet of the ambitwistor string theory to the massless NS-sector background fields was described. In spite of being a worldsheet theory the authors show that conformal invariance persists in the presence of off-shell background fields and the equations of motion for the latter follow from the absence of anomalies in the constraint algebra. For the minimal $U(1)\times U(1)$-gauging the worldsheet approach in \cite{Adamo:2014wea,Adamo:2017nia} and the point particle presented here lead to identical outcomes up to the dimensionality of spacetime which is not constrained for the worldline.  Given that both models have identical spectrum, the compatibility of the field equations is somewhat expected. How this equivalence works in detail is, however, still somewhat mysterious (e.g. \cite{Casali:2016atr}). Indeed the worldline appears to have some more flexibility allowing, in particular, for a cosmological constant as well as the elimination of the dilaton that does not appear to be present in the ambitwistor string. Another approach to recover the field equations of the massless NS-sector in generalized geometry, based on consistent deformations of the graded Poisson structure is described in \cite{Boffo:2019zus}. Finally we would like to stress that our construction of the BRST operator is fully background independent. In string theory, this is an important open problem pioneered by Shatashvili \cite{Shatashvili:1993ps}. 

The rest of this paper is organized as follows. In section \ref{n4sa} we first describe the symmetries of the ${\cal N}=4$ worldline action as well as the (reduced) Hilbert space obtained after gauging some of the $SO(4)$ $R$-symmetry and conclude with the canonical quantization of the constrained system. In section \ref{BRSTQ} we first show that the relative cohomology on the reduced Hilbert space agrees with that before gauging of the $R$-symmetry and describe the complete BV-spectrum of the reduced system. In section \ref{curved} we then couple the system to the different background fields in the spectrum of the theory and derive the equations of motion for the latter from the nilpotency of the BRST charge. We then conclude with suggestions for further work in section \ref{con}. An alternative coupling to the dilaton is presented in the appendix.

\section{$\N=4$ spinning particle and NS-NS spectrum}\label{n4sa}

In \cite{Bonezzi:2018box} the ${\N=4}$ spinning particle was used to provide a worldline description of Einstein gravity. In order to do so, the Hilbert space of the ${\N=4}$ particle had to be constrained by a certain subalgebra of the $so(4)$ $R$-symmetry algebra that  projects the spectrum on the pure gravity sector. Here we will relax such constraints in order to have the full massless NS-NS spectrum of closed strings.

The graded phase space of the model has bosonic coordinates $(x^\mu, p_\mu)$ and fermionic $(\theta^\mu_i, \bar\theta_\nu^j)\,$, where ${\mu=0,...,d-1}$ is a spacetime vector index and $i=1,2$ is a $u(2)$ internal index. Worldline translations and four supersymmetries are generated by the hamiltonian $H$ and supercharges $(q_i, \bar q^i)\,$:
\begin{equation}
H:=p^2\;,\quad q_i:=\theta^\mu_i p_\mu\;,\quad \bar q^i:=\bar\theta^{\mu i}p_\mu\;.    
\end{equation}
In order to describe relativistic massless particles in target space, worldline translations and supersymmetry have to be made local symmetries, as to remove unphysical degrees of freedom and enforce the massless constraint. This is done by introducing a one-dimensional ``supergravity'' multiplet consisting of the einbein $e$ and four gravitinos $(\chi_i, \bar\chi^i)\,$, leading to the worldline action
\begin{equation}\label{action}
S=\int d\tau\,\big[p_\mu\dot x^\mu+i\,\bar\theta_\mu^i\dot\theta_i^\mu-\tfrac{e}{2}\,p^2-i\,\chi_i\,\bar\theta^{\mu i} p_\mu-i\bar\chi^i\,\theta^\mu_i p_\mu\big]\;,
\end{equation}
that is invariant under local reparametrizations\footnote{This form of time reparametrizations is manifestly canonical, being generated by $H$ via Poisson brackets. It differs from the more standard, geometric form, by a trivial transformation.} 
\begin{equation}
\begin{split}
&\delta x^\mu=\xi\,p^\mu\;,\quad\delta p_\mu=0\;,\quad\delta\theta^\mu_i=0\;,\quad\delta\bar\theta^{\mu i}=0\;,\\
&\delta e=\dot\xi\;,\quad\delta\chi_i=0\;,\quad\delta\bar\chi^i=0
\end{split}
\end{equation}
with parameter $\xi(\tau)\,$, and supersymmetries:
\begin{equation}\label{susytrans}
\begin{split}
&\delta x^\mu= i\epsilon_i\,\bar\theta^{\mu i}+i\bar\epsilon^i\,\theta^\mu_i \;,\quad \delta p_\mu=0 \;,\quad\delta \theta^\mu_i= -\epsilon_i\,p^\mu \;,\quad\delta \bar\theta^{\mu i}=-\bar\epsilon^i\,p^\mu \;,\\
&\delta e=2i\,\chi_i\,\bar\epsilon^i+2i\,\bar\chi^i\,\epsilon_i \;,\quad\delta\chi_i=\dot\epsilon_i \;,\quad\delta\bar\chi^i=\dot{\bar\epsilon}^i \;,
\end{split}    
\end{equation}
with odd parameters $\epsilon_i(\tau)$ and $\bar\epsilon^i(\tau)\,$.

Upon canonical quantization, the phase space ``matter'' variables obey the equal time (anti)-commutation relations
\begin{equation}
[x^\mu, p_\nu]=i\,\delta^\mu_\nu\;,\quad\{\theta^\mu_i, \bar\theta_\nu^j\}=\delta^j_i\,\delta^\mu_\nu\;,    
\end{equation}
the other (anti)-commutators being zero. The fermionic system consists thus of $2d$ fermionic oscillators and, by taking the vacuum to obey $\bar\theta_\mu^i\ket{0}=0\,$, a generic state $\ket{\Phi}$ in the Hilbert space can be identified with the wavefunction
\begin{equation}\label{wavefunction}
\Phi(x,\theta_i)=\sum_{m,n=0}^d\phi_{\mu_1...\mu_m| \,\nu_1...\nu_n}(x)\,\theta_1^{\mu_1}...\theta_1^{\mu_m}\,\theta_2^{\nu_1}...\theta_2^{\nu_n}\sim\bigoplus_{m,n}\;m\left\{\,\yng(1,1,1,1)\right.\,\otimes\, n\left\{\yng(1,1,1)\,\right.\;,
\end{equation}
\emph{i.e.} a collection of tensor fields with the symmetries of $(m,n)$ bi-forms, as displayed by the Young diagrams above. 

The quickest way to solve the classical constraints\footnote{The hamiltonian constraint $H=0$ translates as usual to the massless Klein-Gordon equation upon quantization: $\Box\Phi(x,\theta_i)=0\,$.} $q_i=\bar q^i=0$ and to determine the physical spectrum is light-cone quantization: By defining $V^\pm:=\frac{1}{\sqrt2}(V^0\pm V^{d-1})$ for any vector, and assuming $p^+$ to be invertible, one can use the local supersymmetries \eqref{susytrans} to gauge fix $\theta_i^+=\bar\theta^{+i}=0\;$. The constraints $q_i=\bar q^i=0$ are then solved explicitly at the classical level, obtaining $\theta_i^-=\frac{\theta_i^\alpha p_\alpha}{p^+}\,$, where $\alpha=1,...,d-2$ is a transverse index. The same applies for the barred fermions, and the action reduces to 
\begin{equation}
S=\int d\tau\,\big[p_\mu\dot x^\mu+i\,\bar\theta_\alpha^i\dot\theta_i^\alpha-\tfrac{e}{2}\,p^2\big]\;.    
\end{equation}
The only physical oscillators are transverse, thus yielding a unitary spectrum of massless fields
\begin{equation}\label{LCspectrum}
\phi_{\alpha_1...\alpha_m| \,\beta_1...\beta_n}(x)\;,\quad \alpha,\beta=1,...,d-2\;,\quad \Box\phi_{\alpha_1...\alpha_m| \,\beta_1...\beta_n}(x)=0    
\end{equation}
that decompose into irreps of the little group $so(d-2)\,$.

\subsection{Gauging the $R$-symmetries}

The spectrum \eqref{LCspectrum} contains way too many states that, for general $(m,n)$, do not admit minimal coupling to gravity. The relevant NS-NS spectrum we are interested in consists of the level $(1,1)$ field $\phi_{\alpha|\beta}$ that decomposes, as in string theory, into a traceless graviton, a dilaton and Kalb-Ramond two-form:
\begin{equation}
\phi_{\alpha|\beta}=h_{\alpha\beta}+B_{\alpha\beta}+\delta_{\alpha\beta}\,\sigma\;.    
\end{equation}
In order to implement this projection we recall that the action \eqref{action} has a manifest rigid $U(2)$ symmetry that rotates the fundamental (and anti-fundamental) indices of the fermions. In addition, by defining the real fermions $(\Theta^\mu_I, \X_I)\,$, with $I=1,..,4$ as
\begin{equation}
\begin{split}
&\Theta_i^\mu:=\tfrac{1}{\sqrt2}(\theta_i^\mu+\bar\theta^{\mu i})\;,\quad\Theta^\mu_{i+2}:=\tfrac{1}{\sqrt{2}i}(\theta_i^\mu-\bar\theta^{\mu i})\;,\\[2mm]
&\X_i:=\tfrac{1}{\sqrt2}(\chi_i+\bar\chi^{ i})\;,\quad \X_{i+2}:=\tfrac{1}{\sqrt{2}i}(\chi_i-\bar\chi^{i})\;,
\end{split}
\end{equation}
the fermionic action takes the form
\begin{equation}
S_{\rm f}=\int d\tau\,\big[\tfrac{i}{2}\,\Theta_{\mu I}\dot\Theta_I^\mu-i\,\X_I\,\Theta^\mu_I p_\mu\big]\;,    
\end{equation}
that is manifestly invariant under the full $SO(4)$ rigid symmetry generated by the fermion bilinears $J_{IJ}=i\,\Theta_{[I}\cdot\Theta_{J]}\,$. In \cite{Bonezzi:2018box} the maximal set of constraints was imposed on the Hilbert space, in order to project onto the pure gravity sector of the model. On the other hand, the weakest constraint on the Hilbert space compatible with coupling this worldline theory to gravity is obtained by merely gauging a $U(1)\times U(1)$ subgroup of the $R$-symmetry group $SO(4)\,$, that corresponds to combination of level matching and level truncation of a closed string and controls the fermion number eigenvalues $(m,n)$ of \eqref{wavefunction}. More explicitly, we impose the constraints
\begin{equation}
J_i\ket{\Phi}=0\;,\quad J_i:=\theta_i\cdot\bar\theta^i-1\equiv N_i-1\;,\quad\text{ index $i$ not summed.}  \end{equation}
At the level of the worldline action this corresponds to promoting the $U(1)\times U(1)$ subgroup to a local symmetry by introducing two abelian worldline gauge fields\footnote{Here the index $i=1,2$ is a mere label, not a fundamental representation of $U(2)$.} $a_i(\tau)\,$, yielding
\begin{equation}\label{action_gauged}
S=\int d\tau\,\Big[p_\mu\dot x^\mu+i\,\bar\theta_\mu^i\dot\theta_i^\mu-\tfrac{e}{2}\,p^2-i\,\chi_i\,\bar\theta^{\mu i} p_\mu-i\bar\chi^i\,\theta^\mu_i p_\mu-\sum_{i=1,2}a_i\big(\theta_i^\mu\bar\theta_\mu^i-k_i\big)\Big]\;,
\end{equation}
where the one-dimensional Chern-Simons couplings $k_1=k_2=2-\frac{d}{2}$
are required to cancel a quantum ordering effect of the fermions. Choosing ${k_1=m+1-\frac{d}{2}}$ and ${k_2=n+1-\frac{d}{2}}$ projects onto the $(m,n)$ sector of \eqref{wavefunction}. The local supersymmetry transformations of the gravitinos are affected by the $U(1)\times U(1)$ gauging as
\begin{equation}
\delta\chi_i=D_\tau\epsilon_i:=\dot\epsilon_i-ia_i\,\epsilon_i\;,\quad \delta \bar\chi^i=D_\tau\bar\epsilon^i:=\dot{\bar\epsilon}^i+ia_i\,\bar\epsilon^i\;,
\end{equation}
while local $U(1)\times U(1)$ transformations read (here and above the index $i$ is not summed)
\begin{equation}
\delta\theta^\mu_i=i\alpha_i\,\theta^\mu_i\;,\quad \delta\bar\theta_\mu^i=-i\alpha_i\,\bar\theta_\mu^i\;,\quad\delta\chi_i=i\alpha_i\,\chi_i\;,\quad\delta\bar\chi^i=-i\alpha_i\,\bar\chi^i\;,\quad \delta a_i=\dot\alpha_i\;.   
\end{equation}

\subsection{Dirac quantization}

The classical action \eqref{action_gauged} corresponds to the quantum constraint algebra
\begin{equation}\label{u1u1algebra}
\{q_i,\bar q^j\}=\delta_i^j\,H\;,\quad [J_i, q_i]=q_i\;,\quad [J_i,\bar q^i]=-\bar q^i\;,\quad\text{index $i$ not summed,}  
\end{equation}
with the other (anti)-commutators vanishing.
In this description only half of the supercharges can annihilate physical states, while the second half will generate null states, as it is customary in Gupta-Bleuler or old covariant quantization of string theory. The physical state conditions thus read\footnote{Since upon quantization we represent $p_\mu=-i\de_\mu$ on functions of $x\,$, we freely switch between $H$ and $-\Box\,$.}
\begin{equation}
J_i\ket{\Phi}=0\;,\quad \bar q^i\ket{\Phi}=0\;,\quad \Box\ket{\Phi}=0 \;.   
\end{equation}
Solving the $J_i$ constraints one is left with
\begin{equation}
\Phi(x,\theta_i)=\phi_{\mu\vert\nu}(x)\,\theta_1^\mu\theta_2^\nu=\big(\varphi_{\mu\nu}(x)+B_{\mu\nu}(x)\big)\,\theta_1^\mu\theta_2^\nu    
\end{equation}
for $\varphi_{\mu\nu}:=\phi_{(\mu\vert\nu)}$ and $B_{\mu\nu}:=\phi_{[\mu\vert\nu]}$ obeying
\begin{equation}
\Box\varphi_{\mu\nu}=0=\de^\mu\varphi_{\mu\nu}\;,\quad \Box B_{\mu\nu}=0=\de^\mu B_{\mu\nu}\;.    
\end{equation}
The above field equations describe, in a partially gauge-fixed form, a spin two particle and a scalar contained in $\varphi_{\mu\nu}$ together with the two-form $B$. In order to reduce the Lorentz covariant fields to the physical transverse polarizations $B_{\alpha\beta}$ and $\varphi_{\alpha\beta}=h_{\alpha\beta}+\delta_{\alpha\beta}\,\sigma$ 
one has to consider, on top of the above equations, the residual gauge symmetries
\begin{equation}
\delta\varphi_{\mu\nu}=\de_{(\mu}\xi_{\nu)}\;,\quad\Box\xi_\mu=0=\de\cdot\xi\;,\quad \delta B_{\mu\nu}=\de_{[\mu}\lambda_{\nu]}\;,\quad\Box\lambda_\mu=0=\de\cdot\lambda\;,    
\end{equation}
taking care of the trivial transformations $\lambda_\mu=\de_\mu\rho\,$, with $\Box\rho=0\,$,  for the two-form. In the Dirac approach the presence of gauge symmetry manifests with the appearance of null states in the Hilbert space. These are physical states with zero norm and vanishing scalar product with all other physical states, that one can mod out from the physical spectrum. In the present case one can indeed see that fields of the form $\varphi_{\mu\nu}=\de_{(\mu}\xi_{\nu)}$ and $B_{\mu\nu}=\de_{[\mu}\lambda_{\nu]}$ (or equivalently states $\ket{\Phi}=q_i\ket{\Xi^i}$) are null for $\xi_\mu$ and $\lambda_\mu$ transverse and harmonic.

\section{BRST quantization}\label{BRSTQ}

We shall now focus on studying the BRST cohomology associated to the constraint algebra \eqref{u1u1algebra}, as it gives a fully covariant description of the corresponding field theory, with manifest gauge symmetries. We shall do this in two steps. First we show in the following subsection that the BRST chomology of the complete system of constraints obtained by associating ghost-antighost canonical pairs $(c,b)$ to the Hamiltonian and $(C^i, B_i)$ to the $u(1)$'s $J_i$ as well as bosonic superghost pairs $(\bar\gamma^i, \beta_i)$ and $(\gamma_i,\bar\beta^i)$ to the supercharges $q_i$ and $\bar q^i\,$, is equivalent to the relative cohomology of the reduced system obtained by solving the $SO(4)$ constraints. We then continue the analysis within the simpler reduced system. 
\subsection{Reducing the BRST cohomology}
The complete ghost sector satisfies the canonical (anti)-commutation relations
\begin{equation}
\{b,c\}=1\;,\quad\{B_i,C^j\}=\delta^j_i\;,\quad [\beta_i,\bar\gamma^j]=\delta_i^j=[\bar\beta^j,\gamma_i]\;,    
\end{equation}
with ghost numbers
\begin{equation}
{\rm gh}(c,C^i,\gamma_i,\bar\gamma^i)=+1\;,\quad {\rm gh}(b,B_i,\beta_i,\bar\beta^i)=-1    
\end{equation}
The BRST operator associated to the algebra \eqref{u1u1algebra} takes the form
\begin{equation}
\Omega=Q+C^i\J_i \;,   
\end{equation}
where
\begin{equation}
Q:= c\,\Box+\gamma_i\,\bar q^i+\bar\gamma^i\, q_i+\bar\gamma^i\gamma_i\, b \;,\quad Q^2=0   
\end{equation}
is the  BRST operator associated to the gauging of the  ${\N=4}$ supersymmetry alone, and
\begin{equation}
\J_i:=\theta^\mu_i\bar\theta_\mu^i+\gamma_i\bar\beta^i-\bar\gamma^i\beta_i-2\quad\text{($i$ not summed)}    
\end{equation}
are $u(1)\times u(1)$ generators in the matter plus ghost extended space, obeying $[\J_i, Q]=0\,$. With the hermiticity assignments $(\gamma_i)^\dagger=\bar\gamma^i$ and $(\beta_i)^\dagger=-\bar\beta^i\,$, the other ghost variables being self-adjoint, one has
\begin{equation}
Q^\dagger=Q\;,\quad (\J_i)^\dagger=\J_i\quad\longrightarrow\quad \Omega^\dagger=\Omega\;.    
\end{equation}
We choose the ghost vacuum $\ket{0}$ to be annihilated by $(b,B_i,\bar\gamma^i,\bar\beta^i)\,$, so that a general state $\ket{\psi_B}$ in the BRST extended Hilbert space is isomorphic to the wave function\footnote{The functional dependence on the bosonic ghosts $\gamma_i$ and $\beta_i$ is restricted to be polynomial of arbitrary but finite degree.}
$\Psi_B(x,\theta_i\,;c,\gamma_i,\beta_i,C_i)\,$,
on which $(b,B_i,\bar\gamma^i,\bar\beta^i)$ are realized as $(\frac{\de}{\de c}, \frac{\de}{\de C_i},-\frac{\de}{\de \beta_i}, \frac{\de}{\de \gamma_i})\,$. With the given choice of vacuum, the ghost number of the wavefunction is unbounded both from above and below, and the operators $Q$ and $\J_i$ take the form
\begin{equation}
Q=c\,\Box+\gamma_i\,\bar q^i-q_i\,\frac{\de}{\de\beta_i}-\gamma_i\,\frac{\de^2}{\de\beta_i\de c}\;,\quad \J_i=N_{\theta_i}+N_{\gamma_i}+N_{\beta_i}-1=:\N_i-1  \;. \end{equation}

As a first step in the BRST analysis, we will prove that the cohomology of $\Omega$ at ghost number zero is given by the corresponding cohomology of $Q$ on the restricted Hilbert space $\ker\J_1\cap\ker\J_2\,$. To see this, let us first notice that the Hilbert space $\H$ can be decomposed as a double direct sum according to the eigenvalues of the ghost-extended number operators $\N_i$ as
\begin{equation}
\H=\bigoplus_{m,n=0}^\infty\H_{m,n}\;,\quad  \Psi_B=\sum_{m,n=0}^\infty\Psi_{m,n}\;,\quad (\N_1-m)\Psi_{m,n}=0=(\N_2-n)\Psi_{m,n}\;.  
\end{equation}
One further expands the wavefunction according to the $C^i$ dependence as 
\begin{equation}\label{Cexpansion}
\Psi_B=\psi+C^i\,\chi_i+C^1C^2\,\xi\;,\quad (\psi,\chi_i,\xi)=\sum_{m,n=0}^\infty(\psi_{m,n},\chi_{i\,m,n},\xi_{m,n})\;,    
\end{equation}
and similarly the gauge parameter is decomposed as $\Lambda_B=\lambda+C^i\,\eta_i+C^1C^2\,\rho\,$.
The closure relation $\Omega\Psi_B=0$ splits into
\begin{equation}\label{OmegaPsi}
\Omega\Psi_B=0\quad\Rightarrow\quad\left\{
\begin{array}{l}
Q\psi_{m,n}=0\\[1mm] Q\chi_{1\,m,n}=(m-1)\psi_{m,n}\\[1mm]
Q\chi_{2\,m,n}=(n-1)\psi_{m,n}\\[1mm] Q\xi_{m,n}=(n-1)\chi_{1\,m,n}-(m-1)\chi_{2\,m,n}
\end{array}\right.
\end{equation}
as well as the gauge transformations
\begin{equation}\label{deltaPsi}
\delta\Psi_B=\Omega\Lambda_B\quad\Rightarrow\quad\left\{
\begin{array}{l}
\delta\psi_{m,n}=Q\lambda_{m,n}\\[1mm] \delta\chi_{1\,m,n}=-Q\eta_{1\,m,n}+(m-1)\lambda_{m,n}\\[1mm]
\delta\chi_{2\,m,n}=-Q\eta_{2\,m,n}+(n-1)\lambda_{m,n}\\[1mm] 
\delta\xi_{m,n}=Q\rho_{m,n}-(n-1)\eta_{1\,m,n}+(m-1)\eta_{2\,m,n}
\end{array}\right.\;.
\end{equation}
At this point one uses the shift symmetries to gauge away the maximum number of components $\Psi_{m,n}\,$:
\begin{itemize}
\item Use all $\lambda_{m,n}$ with $m\neq1$ to gauge fix $\chi_{1\,m,n}=0$ except $\chi_{1\,1,n}\;\longrightarrow$ closure fixes $\psi_{m,n}=0$ except $\psi_{1,n}$
\item Use all $\eta_{2\,m,n}$ with $m\neq1$ to gauge fix $\xi_{m,n}=0$ except $\xi_{1,n}\;\longrightarrow$ closure fixes $\chi_{2\,m,n}=0$ except $\chi_{2\,1,n}$
\item One is left with the subsystem 
\begin{equation}
\begin{split}
& Q\psi_{1,n}=0\;,\quad Q\chi_{1\,1,n}=0\;,\quad Q\chi_{2\,1,n}=(n-1)\psi_{1,n}\;,\quad Q\xi_{1,n}=(n-1)\chi_{1\,1,n}\;,\\   
& \delta\psi_{1,n}=Q\lambda_{1,n}\;,\quad \delta\chi_{1\,1,n}=-Q\eta_{1\,1,n}\;,\quad \delta\chi_{2\,1,n}=-Q\eta_{2\,1,n}+(n-1)\lambda_{1,n}\;,\\
&\delta\xi_{1,n}=Q\rho_{1,n}-(n-1)\eta_{1\,1,n}
\end{split}    
\end{equation}
\item Repeat the same steps by using the shift symmetries with parameters $(\lambda_{1,n}, \eta_{1\,1,n})$ with $n\neq1$ and the closure relations to further reduce the system to
\begin{equation}\label{4xcohomology}
Q\Psi_a=0\;,\quad \delta\Psi_a=Q\Lambda_a    
\end{equation}
where we grouped $\Psi_a:=(\psi_{1,1},\chi_{i\,1,1},\xi_{1,1})$ and $\Lambda_a:=(\lambda_{1,1},-\eta_{i\,1,1},\rho_{1,1})\,$.
\end{itemize}
Notice that the above components $\Psi_a$ precisely parametrize the subspace $\ker\J_1\cap\ker\J_2\,$, thus proving the above statement. It will now be shown that the non-trivial cohomology at ghost number zero, besides coinciding with the $Q$-cohomology in $\ker\J_1\cap\ker\J_2\,$, is concentrated in the $C^i$-independent part of the wavefunction \eqref{Cexpansion}, \emph{i.e.} in $\psi_{1,1}\,$. We shall drop from now on the subscripts from $(\psi_{1,1},\chi_{i\,1,1},\xi_{1,1})$ by using 
$(\psi,\chi_i,\xi)$ subject to $\J_i(\psi,\chi_i,\xi)=0\,$. The common kernel of the $\J_i$ operators is spanned by the basis elements
\begin{equation}\label{kerkerBasis}
\ker\J_1\cap\ker\J_2={\rm Span}\{(\theta^\mu_1\oplus\gamma_1\oplus\beta_1)\otimes(\theta^\nu_2\oplus\gamma_2\oplus\beta_2)\oplus c\,(\theta^\mu_1\oplus\gamma_1\oplus\beta_1)\otimes(\theta^\nu_2\oplus\gamma_2\oplus\beta_2)\}    
\end{equation}
with the extra four-fold degeneracy $(1\oplus C^i\oplus C^1C^2)$ already taken into account by the decomposition in \eqref{Cexpansion}. The cohomology at ghost number zero, as it can be seen from \eqref{Cexpansion}, is given by the cohomology of $\psi^{(0)}$ together with $\chi_i^{(-1)}$ and $\xi^{(-2)}\,$, where the superscript denotes the ghost number of the corresponding component. We start by considering the $Q$-cohomology of $\xi^{(-2)}$:
\begin{equation}
\J_i\,\xi^{(-2)}=0\;\rightarrow\; \xi^{(-2)}=\rho(x)\,\beta_1\beta_2\;,\qquad
\delta\rho=0\;,\quad \de_\mu\rho=0\;\rightarrow\;\text{trivial}\;.
\end{equation}
Similarly, $\chi_i^{(-1)}$ is shown to contain pure gauge vector fields:
\begin{equation}
\begin{split}
&\J_k\,\chi^{(-1)}_i=0\;\rightarrow\; \chi^{(-1)}_i=v_{i\,\mu}(x)\,\theta_1^\mu\beta_2+\tilde v_{i\,\mu}(x)\,\theta_2^\mu\beta_1+\phi_i(x)\,c\beta_1\beta_2\;,\\[1mm]
& \delta v_{i\,\mu}=\delta\tilde v_{i\,\mu}=i\,\de_\mu\lambda_i\;,\quad \delta\phi_i=\Box\lambda_i\;,\\[1mm]
&\Box v_{i\,\mu}-i\,\de_\mu\phi_i=0=\Box \tilde v_{i\,\mu}-i\,\de_\mu\phi_i\;,\quad \phi_i=-i\,\de\cdot v_i=-i\,\de\cdot \tilde v_i\;,\quad \de_\mu v_{i\,\nu}=\de_\nu \tilde v_{i\,\mu}\;.
\end{split}    
\end{equation}
The scalars are auxiliaries, leaving Maxwell equations for  the four vectors $(v_{i\,\mu}, \tilde v_{i\,\mu})\,$. On the other hand, symmetrizing the last equation one has the Killing equation $\de_{(\mu}v^-_{\nu)\,i}=0$ for $v_i^-:=v_i-\tilde v_i\,$, that does not have acceptable solutions in terms of fluctuating fields, thus yielding $v_i=\tilde v_i\,$. The antisymmetric part of the same equation finally gives $F_{\mu\nu}(v)=0\,$, thus proving that the only non-trivial cohomology at ghost number zero is concentrated in $\psi^{(0)}$ subject to $\J_i\psi^{(0)}=0\,$. We have hence established that the BRST system $\Omega\Psi_B=0\,$, $\delta\Psi_B=\Omega\Lambda_B$ is physically equivalent, at ghost number zero, to the simpler cohomological system
\begin{equation}\label{goodcohomology}
\begin{split}
& Q\,\psi=0\;,\quad \delta\psi=Q\,\Lambda\;,\quad\\
&\J_i\,\psi=0=\J_i\,\Lambda\;,
\end{split}
\end{equation}
where the wavefunction and gauge parameter do not depend on the $C^i$ ghosts: $\psi=\psi(x,\theta_i;c,\gamma_i,\beta_i)\,$.
It should be noticed, however, that the bigger system with charge $\Omega$ effectively generates four copies of the same cohomology\footnote{We thank Maxim Grigoriev for discussions on this point.} (see \eqref{4xcohomology}), since they only differ by a shift in ghost number, that can be anyway redefined. It seems thus better to consider the reduced system \eqref{goodcohomology} as the starting point for the analysis, as well as for the coupling to background fields.

\subsection{Reduced BV-spectrum}
We will now show that the physical states of the above system describe the free propagation of a graviton, a two-form gauge field and a scalar dilaton.
The $\psi$ wavefunction at ghost number zero, obeying $\J_i\psi=0\,$, can be decomposed as
\begin{equation}\label{gh0psi}
\psi^{(0)}=\phi_{\mu\vert\nu}(x)\,\theta^\mu_1\theta^\nu_2+\phi(x)\,\gamma_1\beta_2+\tilde\phi(x)\,\gamma_2\beta_1+A_\mu(x)\,\theta_1^\mu\beta_2c+\tilde A_\mu(x)\,\theta_2^\mu\beta_1c\;. \end{equation}
The gauge symmetry 
\begin{equation}\label{gh0lambda}
\delta\psi^{(0)}=Q\Lambda^{(-1)}\;,\; {\rm with}\quad \Lambda^{(-1)}=\epsilon_\mu\,\theta^\mu_1\beta_2+\tilde\epsilon_\mu\,\theta^\mu_2\beta_1+\eta\,c\beta_1\beta_2
\end{equation}
 reads
\begin{equation}
\begin{split}
&\delta\phi_{\mu\vert\nu}=i\,(\de_\mu\tilde\epsilon_\nu-\de_\nu\epsilon_\mu)\;,\quad \delta A_\mu=i\,\de_\mu\eta-\Box\epsilon_\mu\;,\quad \delta \tilde A_\mu=i\,\de_\mu\eta-\Box\tilde\epsilon_\mu\;,\\
&\delta\phi=-i\,\de\cdot\epsilon-\eta\;,\quad \delta\tilde \phi=-i\,\de\cdot\tilde\epsilon-\eta \;.
\end{split}    
\end{equation}
Upon the field redefinitions
\begin{equation}
\begin{split}
&\varphi_{\mu\nu}:=\phi_{(\mu\vert\nu)}\;,\quad B_{\mu\nu}:=\phi_{[\mu\vert\nu]}\;,\quad i\,A_\mu^\pm:=\tilde A_\mu\pm A_\mu\;,\quad \phi^\pm:=\tilde\phi\pm\phi\;,\\
&\varepsilon_\mu:=i(\tilde\epsilon_\mu-\epsilon_\mu)\;,\quad \lambda_\mu:=i(\tilde\epsilon_\mu+\epsilon_\mu)
\end{split}
\end{equation}
the closure equations become
\begin{equation}
\begin{split}
&\Box\varphi_{\mu\nu}-\de_{(\mu}A^-_{\nu)}=0\;,\quad \Box B_{\mu\nu}-\de_{[\mu}A^+_{\nu]}=0\;,\\
& A^-_\mu=2\,\de\cdot\varphi_\mu+\de_\mu\phi^-\;,\quad A^+_\mu=2\,\de^\lambda B_{\lambda\mu}-\de_\mu\phi^+\;,\\
& \Box\phi^\pm+\de\cdot A^\pm=0
\end{split}    
\end{equation}
with gauge symmetries
\begin{equation}
\begin{split}
& \delta\varphi_{\mu\nu}=\de_{(\mu}\varepsilon_{\nu)}\;,\quad\delta B_{\mu\nu}=\de_{[\mu}\lambda_{\nu]}\;,\\
& \delta A^-_\mu=\Box\varepsilon_\mu\;,\quad\delta A^+_\mu=\Box\lambda_\mu+2\,\de_\mu\eta\;,\\
& \delta\phi^-=-\de\cdot\varepsilon\;,\quad \delta\phi^+=-\de\cdot\lambda-2\,\eta\;.
\end{split}    
\end{equation}
By solving for  the auxiliary vectors $A_\mu^\pm$ one obtains the system
\begin{equation}
\begin{split}
&\Box\varphi_{\mu\nu}-2\,\de_{(\mu}\de\cdot\varphi_{\nu)}-\de_\mu\de_\nu\phi^-=0\;,\quad \Box\phi^-+\de\cdot\de\cdot\varphi=0\;,\\    
& \Box B_{\mu\nu}+2\,\de_{[\mu}\de^\lambda B_{\nu]\lambda}\equiv\de^\lambda H_{\lambda\mu\nu}(B)=0
\end{split}    
\end{equation}
where the scalar $\phi^+$ has dropped out\footnote{It can also be gauged away by the shift symmetry $\eta$.}. The scalar $\phi^-$ on the other hand is mixed with the trace of $\varphi_{\mu\nu}\,$, the gauge invariant combination being $\sigma:=\phi^-+\varphi^\lambda{}_\lambda\,$, in terms of which the spin two -- spin zero system becomes
\begin{equation}
\Box\varphi_{\mu\nu}-2\,\de_{(\mu}\de\cdot\varphi_{\nu)}+\de_\mu\de_\nu\varphi^\lambda{}_\lambda  =\de_\mu\de_\nu\sigma\;,\quad \Box\sigma=0\;.   
\end{equation}
 The above system, together with the two-form $B_{\mu\nu}\,$, coincides with the linearized field equations of the massless NS-NS sector of closed strings\footnote{Notice that $\Box\varphi_{\mu\nu}-2\,\de_{(\mu}\de\cdot\varphi_{\nu)}+\de_\mu\de_\nu\varphi^\lambda{}_\lambda=-2\,R_{\mu\nu}^{\rm lin}(\eta+\varphi)\,$. }. The fluctuation $\varphi_{\mu\nu}$ is the string-frame graviton, the Einstein-frame one being given by
\begin{equation}
h_{\mu\nu}:=\varphi_{\mu\nu}-\tfrac{1}{d-2}\,\eta_{\mu\nu}\,\sigma\;,   
\end{equation}
for which one has the decoupled free equations
\begin{equation}
 \Box h_{\mu\nu}-2\,\de_{(\mu}\de\cdot h_{\nu)}+\de_\mu\de_\nu h^\lambda{}_\lambda=0\;,\quad \Box\sigma=0\;.   
\end{equation}
The cohomological system \eqref{goodcohomology} does not only provide field equations and gauge symmetries for the physical fields, but it also encodes the spacetime BV spectrum and BRST transformations \cite{Barnich:2003wj,Barnich:2004cr}. Similarly to string field theory, one can assign spacetime ghost number and parity to the component fields of $\psi(x,\theta_i;c,\gamma_i,\beta_i)$ by demanding $\psi$ to have total even parity and ghost number zero, interpreting it as a ``string field''. Explicitly, the most general state obeying $\J_i\psi=0$ can be decomposed as
\begin{equation}\label{BV superfield}
\begin{split}
{\rm ker}\J_i\ni\psi &= (\varphi_{\mu\nu}+B_{\mu\nu})\,\theta^\mu_1\theta^\nu_2+\phi\,\gamma_1\beta_2+\tilde\phi\,\gamma_2\beta_1+A_\mu\,\theta_1^\mu\beta_2c+\tilde A_\mu\,\theta_2^\mu\beta_1c\\[1mm]
& +\xi_\mu\,\theta_1^\mu\beta_2+\tilde\xi_\mu\,\theta_2^\mu\beta_1+\eta\,\beta_1\beta_2c+\lambda\,\beta_1\beta_2\\[2mm]
& +(\varphi^*_{\mu\nu}+B^*_{\mu\nu})\,\theta^\mu_1\theta^\nu_2c+\phi^*\,\gamma_1\beta_2c+\tilde\phi^*\,\gamma_2\beta_1c+A^*_\mu\,\theta_1^\mu\gamma_2+\tilde A^*_\mu\,\theta_2^\mu\gamma_1\\[1mm]
& +\xi^*_\mu\,\theta_1^\mu\gamma_2c+\tilde\xi^*_\mu\,\theta_2^\mu\gamma_1c+\eta^*\,\gamma_1\gamma_2+\lambda^*\,\gamma_1\gamma_2c\;.
\end{split}
\end{equation}
 The first line above contains the fields at ghost number zero displayed in \eqref{gh0psi}, namely the graviton and Kalb-Ramond two-form, the dilaton and a pure gauge scalar,  contained in $\phi\pm\tilde\phi$ respectively,  and the two auxiliary vectors $A_\mu\pm\tilde A_\mu$ associated to longitudinal modes of $\varphi_{\mu\nu}$ and $B_{\mu\nu}\,$. The second line contains the vector ghosts $\xi_\mu\pm\tilde\xi_\mu$ associated to the spin two and two-form gauge symmetries, the ghost for ghost $\lambda$ corresponding to the reducibility of the two-form gauge symmetry, and the scalar ghost $\eta$ for the spin one gauge transformation of $A_\mu+\tilde A_\mu\,$. The second half displays all the corresponding antifields, thus yielding the minimal BV spectrum plus auxiliaries.
 The BV-extended gauge symmetry is given by $\delta\psi=Q\Lambda\,$, where the gauge parameter string field $\Lambda$ is assigned total odd parity and ghost number $-1\,$, while the spacetime BRST differential acts as $s\,\psi=Q\psi\,$ (see for instance \cite{Grigoriev:2006tt}).

\section{${\cal N}=4$ point particle coupled to background fields}\label{curved}

In this section we discuss the coupling of our worldline model to background fields including the metric $g_{\mu\nu}\,$, the Kalb-Ramond field $\B_{\mu\nu}$ as well as the dilaton $\Phi$,\footnote{Here we use the symbols $g_{\mu\nu}\,$, $\B_{\mu\nu}$ and $\Phi$ for background fields in order to distinguish them from the corresponding states in the Hilbert space, denoted by $\varphi_{\mu\nu}\,$, (or $h_{\mu\nu}\,$,) $B_{\mu\nu}$ and $\sigma\,$ .} taking the reduced cohomological system \eqref{goodcohomology} as a starting point for the deformation. As in \cite{Bonezzi:2018box} we take the fermions with flat Lorentz indices, \emph{i.e.}  $(\theta_i^a, \bar\theta^{i\,a})$ together with a background vielbein $e_\mu^a(x)$ and spin connection that is torsion-free. 
\subsection{Pure gravity}\label{qgc}
The coupling to a background metric has been described in detail in \cite{Bonezzi:2018box} in terms of covariant derivative operators
\begin{equation}
\hat\nabla_\mu:=\de_\mu+\omega_{\mu\, ab}\,\theta^a\!\cdot\bar\theta^b\;,\quad\quad [\hat\nabla_\mu, \hat\nabla_\nu]=R_{\mu\nu\lambda\sigma}\,\theta^\lambda\!\cdot\bar\theta^\sigma=:\bm{R}_{\mu\nu} \;, \end{equation}
where $\theta^\mu_i:=e^\mu_a(x)\,\theta^a_i\,$, and similarly for $\bar\theta^{i\,\mu}\,$. The difference in the present treatment is that we consider a bigger Hilbert space, defined by ${\rm ker}\J_i$, compared to \cite{Bonezzi:2018box} where all of $SO(4)$ was gauged.  The curved space supercharges and Laplacian
\begin{equation}\label{Csupercharges}
q_i:=-i\,\theta_i^a\,e^\mu_a\,\hat\nabla_\mu\;,\quad \bar q^i:=-i\,\bar\theta^{i\,a}\,e^\mu_a\,\hat\nabla_\mu\;,\quad \nabla^2:=g^{\mu\nu}\hat\nabla_\mu \hat\nabla_\nu-g^{\mu\nu}\,\Gamma^\lambda_{\mu\nu}\,\hat\nabla_\lambda\equiv \frac{1}{\sqrt g}\hat\nabla_\mu\sqrt{g}g^{\mu\nu}\hat\nabla_\nu    
\end{equation}
have commutation relations
\begin{equation}
\begin{split}
&\{q_i, q_j\}=-\theta^\mu_i\theta^\nu_j\, \bm{R}_{\mu\nu}\;,\quad \{\bar q^i, \bar q^j\}=-\bar\theta^{\mu\,i}\bar\theta^{\nu\,j}\,\bm{R}_{\mu\nu}\;,\quad \{q_i, \bar q^j\}=-\delta_i^j\,\nabla^2-\theta^\mu_i\bar\theta^{\nu\,j}\,\bm{R}_{\mu\nu}\;,\\[3mm]
& [\nabla^2, q_i]=i\,\theta^\mu_i\big(2\,\bm{R}_{\mu\nu}\,\hat\nabla^\nu-\nabla^\lambda\bm{R}_{\lambda\mu}-R_{\mu\nu}\hat\nabla^\nu\big)\;,\\[3mm]
& [\nabla^2, \bar q^i]=i\,\bar\theta^{\mu\,i}\big(2\,\bm{R}_{\mu\nu}\,\hat\nabla^\nu-\nabla^\lambda\bm{R}_{\lambda\mu}-R_{\mu\nu}\hat\nabla^\nu\big)\;,
\end{split}    
\end{equation}
where $\nabla^\mu\bm{R}_{\mu\nu}:=(\nabla^\mu R_{\mu\nu\lambda\sigma})\theta^\lambda\!\cdot\bar\theta^\sigma\,$.
The corresponding BRST operator $Q$ is given by 
\begin{equation}\label{curvedQ}
\begin{split}
& Q=c\,\triangle-i\,S^\mu\hat\nabla_\mu+\bar\gamma^i\gamma_i\,b\;,\quad\triangle:=\nabla^2+\Re\,;\qquad S^\mu:=\bar\gamma^i\theta^\mu_i+\gamma_i\,\bar\theta^{\mu\,i}\;.
\end{split}    
\end{equation}
where $\Re=R_{\mu\nu\lambda\sigma}\,\theta^\mu\!\cdot\bar\theta^\nu\,\theta^\lambda\!\cdot\bar\theta^\sigma$ is a non-minimal coupling \cite{Bonezzi:2018box} needed for $Q$ to be nilpotent. From 
\begin{equation}\label{nilpotent}
Q^2=\bar\gamma\!\cdot\gamma\,\triangle-(S^\mu\hat\nabla_\mu)^2-ic\,[\triangle, S^\mu\hat\nabla_\mu]\;,   
\end{equation}
one finds that the two independent obstructions to nilpotency of $Q$ read 
\begin{equation}\label{first obstruction}
\begin{split}
 \bar\gamma\!\cdot\gamma\,\triangle-(S^\mu\hat\nabla_\mu)^2 &= -\tfrac12\,S^\mu S^\nu\,\bm{R}_{\mu\nu}+\bar\gamma\!\cdot\gamma\,\Re\\[2mm]
& \!\!\!\stackrel{{\rm ker}\J_i}{=} \bar\gamma\!\cdot\theta^\mu\,\gamma\!\cdot\bar\theta^\nu\,R_{\mu\nu}+\bar\gamma\!\cdot\gamma\,R_{\mu\nu}\,\theta^\mu\!\cdot\bar\theta^\nu\;,    
\end{split}    
\end{equation}
and 
\begin{equation}\label{second obstruction}
\begin{split}
[\triangle,S^\mu\hat\nabla_\mu] &= S^\mu\nabla^\lambda\bm{R}_{\lambda
\mu}- S^\mu\nabla_\mu\Re\\[2mm]
& \!\!\!\stackrel{{\rm ker}\J_i}{=} S^\mu\nabla^\lambda\bm{R}_{\lambda
\mu}- \big(2 \nabla^\lambda\bm{R}_{\lambda
\mu}\gamma\!\cdot\bar\theta^\mu-S^\mu\nabla_\mu R_{\nu\lambda}\theta^\nu\!\cdot\bar\theta^\lambda\big)\;.
\end{split}    
\end{equation}
In the second lines the above obstructions have been evaluated on the restricted Hilbert space ${\rm ker}\J_i\,$. Upon normal ordering, \emph{i.e.} moving all barred oscillators to the right, this amounts to setting to zero any contribution with at least three barred oscillators. The above result shows that $Q^2=0$ on Ricci-flat backgrounds as in string theory (upon recalling that on a Ricci-flat manifold one has $\nabla^\mu R_{\mu\nu\lambda\sigma}=0$). As explained in \cite{Bonezzi:2018box}, it is possible to turn on an Einstein background with non-vanishing cosmological constant of any sign at the price of restricting further the Hilbert space, which amounts to projecting away the $B$-field.

\subsection{Coupling the $B$-field}
In order to  additionally couple an external background field $\B_{\mu\nu}$, we consider the deformed covariant derivative 
\begin{equation} \label{covarientD_with_B}
\hat{D}_\mu:=\partial_\mu + \omega_{\mu a b}\, \theta^a \!\cdot \bar{\theta}^b + H_{\mu a b}\left(\theta^a_2 \bar{\theta}^{2b}-\theta^a_1 \bar{\theta}^{1b}\right)\;,  
\end{equation}
where $H_{\mu\nu\lambda}$ is the field strength of $\B_{\mu\nu}$. Note that the term that multiplies $H_{\mu a b}$ in (\ref{covarientD_with_B}) breaks the $R$-symmetry\footnote{This is a manifestation of the fact that in string theory the Kalb-Ramond field couples to the left- and right-moving fermions with opposite sign.} down to the subgroup $U(1)\times U(1)$.  The curved space supercharges are defined in the same way as in section \ref{qgc}. 
To begin with, let us introduce the notation for twisted variables:
\begin{equation}
\vartheta^a_i:=(-1)^i\,\theta^a_i=(-\theta^a_1, \theta^a_2)\;,\quad \bar\vartheta^{a\,i}:=(-1)^i\,\bar\theta^{a\,i}=(-\bar\theta^{a\,1}, \bar\theta^{a\,2}) \;.   
\end{equation}
We then define Lorentz generators $S^{ab}$ and twisted generators $T^{ab}$
\begin{equation}
S^{ab}:=2\,\theta^{[a}\!\cdot\bar\theta^{b]} \;,\quad T^{ab}:=2\,\theta^{[a}\!\cdot\bar\vartheta^{b]}\equiv2\,\vartheta^{[a}\!\cdot\bar\theta^{b]}=2\,(\theta^{[a}_2\bar\theta^{b]2}-\theta_1^{[a}\bar\theta^{b]1})   
\end{equation}
that obey the extended $so(d)\oplus so(d)$ algebra\footnote{The two commuting $so(d)$ algebras are given by $S^{ab}_i:=2\,\theta^{[a}_i\bar\theta^{b]i}$ with $i=1,2$ not summed.}
\begin{equation}
[S^{ab}, S^{cd}]=4\,\eta^{[c[b}S^{a]d]}\;,\quad [S^{ab}, T^{cd}]=4\,\eta^{[c[b}T^{a]d]}\;,\quad [T^{ab}, T^{cd}]=4\,\eta^{[c[b}S^{a]d]}\;.  
\end{equation}
The generalized covariant derivative operators can be recast in the form
\begin{equation}
\hat D_\mu:=\de_\mu+\tfrac12\,\omega_{\mu ab}\,S^{ab}+\tfrac12\,H_{\mu ab}\,T^{ab}= \hat\nabla_\mu+\tfrac12\,H_{\mu ab}\,T^{ab}=\de_\mu+\omega_{\mu ab}\,\theta^a\!\cdot\bar\theta^b+H_{\mu ab}\,\theta^a\!\cdot\bar\vartheta^b   
\end{equation}
with $H_{\mu\nu\lambda}:=3\,\de_{[\mu}\B_{\nu\lambda]}\,$. 
Tensors in the ${\N=4}$ Hilbert space have the form
\begin{equation}
t_{\mu[m]|\nu[n]}\sim m\left\{\,\yng(1,1,1,1)\right.\,\otimes\, n\left\{\yng(1,1,1)\,\right.    
\end{equation}
on which the operator $\hat D_\mu$ acts as
\begin{equation}
D_\mu t_{\nu[n]|\lambda[m]}=\nabla_\mu t_{\nu[n]|\lambda[m]}-n\,H_{\mu\nu}{}^\alpha t_{\alpha\nu[n-1]|\lambda[m]}+m\,H_{\mu\lambda}{}^\alpha t_{\nu[n]|\alpha\lambda[m-1]}\;, \end{equation}
and has the hermiticity property $\hat D_\mu^\dagger=-(\hat D_\mu+\Gamma_{\mu\lambda}^\lambda)$ with respect to the inner product
\begin{equation}
\left\langle V,W\right\rangle=\int d^dx\sqrt{g}\,V_{\mu[m]|\nu[n]}\,W^{\mu[m]|\nu[n]}   \;. 
\end{equation}
The commutator of covariant derivatives yields
\begin{equation}
\bm{C}_{\mu\nu}:=[\hat D_\mu, \hat D_\nu]=\tfrac12\,\R_{\mu\nu ab}\, S^{ab}+\nabla_{[\mu}H_{\nu]ab}\,T^{ab}  \;,
\end{equation}
where we defined the generalized Riemann tensor
\begin{equation}
\R_{\mu\nu\lambda\sigma}:=R_{\mu\nu\lambda\sigma}-H_{\mu\lambda}{}^\alpha H_{\nu\sigma\alpha}+H_{\nu\lambda}{}^\alpha H_{\mu\sigma\alpha}\;,
\end{equation}
that obeys
\begin{equation}
\R_{\mu\nu\lambda\sigma}=\R_{[\mu\nu]\lambda\sigma}=\R_{\mu\nu[\lambda\sigma]}=\R_{\lambda\sigma\mu\nu}\;,
\end{equation}
and thus admits a single and symmetric generalized Ricci tensor
\begin{equation}
\R_{\mu\nu}:=\R^\lambda{}_{\mu\lambda\nu}=R_{\mu\nu}-H_\mu{}^{\lambda\sigma}H_{\nu\lambda\sigma}\;.
\end{equation}
However, it does not satisfy the Bianchi identity:
\begin{equation}
\R_{[\mu\nu\lambda]\sigma}=2\,H_{[\mu\nu}{}^\alpha H_{\lambda]\sigma\alpha}\;.
\end{equation}
The supercharges are defined as
\begin{equation}
q_i:=-i\,\theta^a_i\,e^\mu_a\,\hat D_\mu=-i\,\theta^\mu_i\,\hat D_\mu\;,\quad \bar q^i:=-i\,\bar\theta^{a\,i}\,e^\mu_a\,\hat D_\mu=-i\,\bar\theta^{\mu\,i}\,\hat D_\mu\;,
\end{equation}
and obey the algebra
\begin{equation}
\begin{split}
&\{q_i,q_j\}=-\theta^\mu_i\theta^\nu_j\,\bm{C}_{\mu\nu}-(\theta^\mu_i\vartheta^\nu_j+\vartheta^\mu_i\theta^\nu_j)H_{\mu\nu}{}^\lambda\hat D_\lambda\;,\\[2mm]
&\{\bar q^i,\bar q^j\}=-\bar\theta^{\mu\,i}\bar\theta^{\nu\,j}\,\bm{C}_{\mu\nu}-(\bar\theta^{\mu\,i}\bar\vartheta^{\nu\,j}+\bar\vartheta^{\mu\,i}\bar\theta^{\nu\,j})H_{\mu\nu}{}^\lambda\hat D_\lambda\;,\\[2mm]
&\{q_i,\bar q^j\}=-\delta_i^j\,D^2-\theta^\mu_i\bar\theta^{\nu\,j}\,\bm{C}_{\mu\nu}-(\theta^\mu_i\bar\vartheta^{\nu\,j}+\vartheta^\mu_i\bar\theta^{\nu\,j})H_{\mu\nu}{}^\lambda\hat D_\lambda\;,
\end{split}    
\end{equation}
with the generalized Laplacian defined by
\begin{equation}
D^2:=g^{\mu\nu}\hat D_\mu\hat D_\nu-g^{\mu\nu}\Gamma^\lambda_{\mu\nu}\hat D_\lambda = \frac{1}{\sqrt g}\hat D_\mu\sqrt{g}g^{\mu\nu}\hat D_\nu\;,\quad (D^2)^\dagger=D^2\;.    
\end{equation}
In analogy with the coupling to gravity in the last subsection, we make the following Ansatz for the BRST charge (we recall that $S^\mu=\bar\gamma\cdot\theta^\mu+\gamma\cdot\bar\theta^\mu$):
\begin{equation}\label{QwithB}
Q=c\,\triangle-i\,S^\mu\hat D_\mu+\bar\gamma\cdot\gamma\,b\;,\quad\triangle:=D^2+\Re\;,
\end{equation}
where $\Re$ contains possible non-minimal couplings to be determined by insisting on nilpotency. One obstruction to the nilpotency of $Q$ is given by
\begin{equation}\label{first obstruction with B}
\begin{split}
\bar\gamma\cdot\gamma\,\triangle-(S^\mu\hat D_\mu)^2&=-\tfrac12\,S^\mu S^\nu\,\bm{C}_{\mu\nu}-S^\mu T^\nu\,H_{\mu\nu}{}^\lambda\hat D_\lambda+\bar\gamma\cdot\gamma\,\Re \\[2mm]
&\!\!\!\stackrel{\ker\J_i}{=} \bar\gamma\cdot\theta^\mu\gamma\cdot\bar\theta^\nu\,\R_{\mu\nu}-\bar\gamma\cdot\theta^\mu\gamma\cdot\bar\vartheta^\nu\,\nabla^\lambda H_{\lambda\mu\nu}+\bar\gamma\cdot\gamma\,\Re\rvert_{\ker \J_i}\;.
\end{split}    
\end{equation}
To make it vanish, one has to impose
\begin{equation}\label{BGfieldequations}
\R_{\mu\nu}=R_{\mu\nu}-H_\mu{}^{\lambda\sigma}H_{\nu\lambda\sigma}=0\;,\quad \nabla^\lambda H_{\lambda\mu\nu}=0\;, \end{equation}
that are precisely the (two derivative) field equations for the massless NS-NS sector of closed strings, in case of a constant dilaton background. Notice that consistency of the field equation $R_{\mu\nu}=H_\mu{}^{\lambda\sigma}H_{\nu\lambda\sigma}$ with the Bianchi identities of the Ricci tensor and the $H$ field strength requires $\nabla_\mu H^2=0\,$. This is in agreement with closed string field equations, as a constant dilaton background requires $H^2=0\,$. An additional requirement for \eqref{first obstruction with B} to vanish is that the non-minimal coupling $\Re$ be taken as to obey $\bar\gamma\cdot\gamma\,\Re\vert_{\ker\J_i}=0\,$.

The second obstruction comes from
\begin{equation}\label{D2DobsB}
\begin{split}
[\triangle, S^\mu\hat D_\mu]&=\R_{\mu\nu}\,S^\mu\hat D^\nu+\nabla^\lambda H_{\lambda\mu\nu}\,T^\mu\hat D^\nu-H^{\mu\nu}{}_\lambda\,T^\lambda\bm{C}_{\mu\nu}-2\,S^\mu\,\bm{ C}_{\mu\nu}\hat D^\nu\\[2mm]
&+S^\nu\,\nabla^\mu\bm{C}_{\mu\nu}+S^\mu\,H^{\nu\lambda}{}_\rho\big[2\,\nabla_{[\mu} H_{\nu]\lambda\sigma}\,S^{\rho\sigma}+\R_{\mu\nu\lambda\sigma}\,T^{\rho\sigma}\big]+[\Re,S^\mu\hat D_\mu]\;,
\end{split}    
\end{equation}
where we defined a second ghost-valued vector
\begin{equation}
 T^\mu:= \bar\gamma\cdot\vartheta^\mu+\gamma\cdot\bar\vartheta^\mu\,.   
\end{equation}
In order to evaluate the obstruction on the constrained Hilbert space $\ker\J_i\,$, one may use the identity
\begin{equation}
T^a S^{bc}\stackrel{\ker\J_i}{=}-S^a T^{bc}+4\,\eta^{a[b}\gamma\cdot\bar\vartheta^{c]}\;,\quad T^aT^{bc}\stackrel{\ker\J_i}{=}-S^aS^{bc}+4\,\eta^{a[b}\gamma\cdot\bar\theta^{c]}    \;,
\end{equation}
to relate seemingly different tensor structures.

It turns out that, upon evaluating \eqref{D2DobsB} on ${\rm ker}\J_i\,$, the term $S^\mu\,\bm{C}_{\mu\nu}\hat D^\nu$ is the only one that explicitly needs to be canceled by a contribution from $\Re\,$. The only way to make the obstruction $[\triangle, S^\mu\hat D_\mu]$ vanish is thus to choose the non-minimal coupling $\Re$ proportional to $S^{\mu\nu}\bm{C}_{\mu\nu}\,$, that complies with the requirement $\bar\gamma\cdot\gamma\,\Re\vert_{\ker\J_i}=0$ on-shell. We thus choose
\begin{equation}\label{full hamiltonian}
\triangle=D^2+\tfrac14\,(S^{\mu\nu}\bm{C}_{\mu\nu}+\bm{C}_{\mu\nu}S^{\mu\nu})=D^2+\tfrac14\,\R_{\mu\nu\lambda\sigma}\,S^{\mu\nu}S^{\lambda\sigma}+\tfrac12\,\nabla_\mu H_{\nu\lambda\sigma}\,S^{\mu\nu}T^{\lambda\sigma}-\tfrac12\,\nabla^\lambda H_{\lambda\mu\nu}\,T^{\mu\nu}\;,
\end{equation}
where we wrote the non-minimal coupling in a manifestly hermitean form  
\begin{equation}
\begin{split}
\Re=\tfrac14\,(S^{\mu\nu}\bm{C}_{\mu\nu}+\bm{C}_{\mu\nu}S^{\mu\nu})\equiv\tfrac12\,S^{\mu\nu}\bm{C}_{\mu\nu}-\tfrac12\,\nabla^\lambda H_{\lambda\mu\nu}\,T^{\mu\nu} \,,
\end{split}    
\end{equation}
With this choice we finally get

\begin{eqnarray}
 [ {\triangle ,{S^\mu }{{\hat D}_\mu }} ]& \stackrel{{\rm ker}\J_i}{=} &  {\nabla_
 \mu}{\cal R}{\,_{
\nu\rho }} (S_-^\rho S^{\mu\nu}+S^\mu\theta^\nu\!\cdot\bar\theta^\rho)+ {H_{\rho \lambda\mu}} {{\nabla_\sigma }H_{\,\,\,\,\,\,\,\,\,\nu}^{\sigma\lambda }}\,S_-^\rho {S^{\mu\nu}} 
  - \nabla_\mu \nabla^\lambda H_{\lambda\rho\nu}\, S_-^\rho T^{\mu\nu} \nonumber\\ &&
 +{\cal R}{\,_{\mu \lambda}}H^\lambda{}_{\rho\nu}\,\big(2\,S^\rho{\theta ^{(\mu}}\! \cdot {{\bar \vartheta }^{\nu)}}-S^\rho_-T^{\mu\nu}\big)\\ 
  &&+ 2\,
  \bar\gamma\cdot\theta^
  \mu\left( \tfrac13\,{\nabla_\mu}H^2 -H_\mu{}^{\nu\lambda}\,\nabla^\rho H_{\rho\nu\lambda}  -\nabla^\nu\R_{\mu\nu}\right)+\nabla^\lambda H_{\lambda\mu\nu}\,T^\mu_-\hat D^\nu\;, \nonumber
\end{eqnarray}
where
\begin{equation}
    S_-^\mu:=\bar\gamma\cdot\theta^\mu-\gamma\cdot\bar\theta^\mu\,,\quad T^\mu_-:= \bar\gamma\cdot\vartheta^\mu-\gamma\cdot\bar\vartheta^\mu\,.  
\end{equation}
Thus the obstruction vanishes on-shell assuming the equation of motion  (\ref{BGfieldequations}). Summarizing,
\begin{equation}
Q^2\stackrel{\ker\J_i}{=}0\quad{\rm for}\quad R_{\mu\nu}=H_\mu{}^{\lambda\sigma}H_{\nu\lambda\sigma}\;,\quad \nabla^\lambda H_{\lambda\mu\nu}=0    
\end{equation}
with $Q$ as in \eqref{QwithB} and $\triangle$ as in  \eqref{full hamiltonian}. 

\subsubsection{Vertex operator for the $B$-field}
In order to construct the verterx operator for the $B$-field we proceed as in the case of pure gravity in \cite{Bonezzi:2018box}. Namely, we write 
\begin{align}
    V=Q-Q_0 = V_h + V_{b}\,,
\end{align}
where $Q$ is the BRST charge with infinitesimal background field perturbations $g_{\mu\nu}=\eta_{\mu\nu}+h_{\mu\nu}$ and $\B_{\mu\nu}=b_{\mu\nu}$, $Q_0$ is the BRST charge with trivial background and $V_h$ and $V_{b}$ are the vertex operators for the graviton and the $B$-field respectively. The vertex operator $V_{b}$ splits in two parts, according to the $c$-ghost:
\begin{align}
    V_{b}=c\,W_I+W_{II}\,.
\end{align}
 Taking $b_{\mu\nu}(x)= b_{\mu\nu}\,e^{ikx}\,$, where the polarization obeys $b_{\mu\nu} = - b_{\nu\mu}$ and $b_\mu \cdot k = 0$\,, we get
\begin{align}
    W_{\rm I} &= \left(i\left(b_{\mu\nu}k^\lambda-2\,b_\mu^{ \ \lambda}k_\nu\right)\partial_\lambda + b_{\mu\nu}k^2 +b_{\lambda\mu}k_{\nu}k_{\rho}\,S^{\lambda\rho}\right)
    \,T^{\mu\nu}\,e^{ikx}, \\
    W_{\rm II} &= 
    \tfrac12 S^\lambda\left(b_{\mu\nu}k_\lambda+ 2\,b_{\lambda \mu}k_\nu\right)\,T^{\mu\nu}\,e^{ikx}.
\end{align}
The actual one-particle state for $\B$-excitations is obtained as in \cite{Bonezzi:2018box}, by acting with $W_{\rm II}$ on a particular diffeomorphism ghost state in (\ref{BV superfield}),
\begin{align}
    W_{\rm II}\,\xi_\rho\, \beta_{[1}\theta^\rho_{2]} \left|0\right>  = b_{\mu\nu}\,e^{ikx}\,\theta^\mu_1\theta^\nu_2\left|0\right> ,
\end{align}
where $\xi_\rho$ is chosen such that $\xi^\mu b_{\mu\nu} = 0$ and $\xi^\mu k_\mu=-1$. As a consistency check we can calculate the two graviton and one $B$-field scattering amplitude. Repeating the procedure outlined in \cite{Bonezzi:2018box} one finds
\begin{align}
    \left<h^{(2)}\right|V_b\left|h^{(1)}\right> = \left<\xi^{(2)}\right|V_h^{(2)}\,c\,W_{\rm I}\,V_h^{(1)}\left|\xi^{(1)}\right>=0\,,
\end{align}
as it should be. The scattering amplitude for two $B$-fields with polarizations $b$ and $b^{(1)}$, and one graviton is, in turn,
\begin{align}
    \left<b \vphantom{b^{(0)}}\right|V^{(2)}_h\left|b^{(1)}\right> &= -4\,{\rm Tr}\left(b\cdot b^{(1)}\right)k^{(1)}\cdot \epsilon^{(2)} \cdot k^{(1)} + 4k^{(2)} \cdot k^{(2)} \ {\rm Tr}\left(b^{(1)}\cdot b\cdot \epsilon^{(2)}\right) \nonumber\\
    &+ 8\left(k^{(1)}\cdot \epsilon^{(2)} \cdot b\cdot b^{(1)}\cdot k^{(2)} - k^{(2)}\cdot  b\cdot b^{(1)}\cdot \epsilon^{(2)} \cdot k^{(1)}- k^{(2)}\cdot  b\cdot  \epsilon^{(2)} \cdot b^{(1)} \cdot k^{(2)} \right),
\end{align}
with $k^{(1)}+k^{(2)}=k$. Similarly, one can consider one $B$-field in the vertex operator and the other one, as well as the graviton, in the external bra and ket states:
\begin{align}\label{cfft}
    \left<h^{(2)}\right|V_b\left|b^{(1)}\right> &= 4\, {\rm Tr}\left(\epsilon^{(2)}\cdot b^{(1)} \cdot b\right)k^2 + 8 \,{\rm Tr}\left(\epsilon^{(2)}\cdot b^{(1)} \cdot b\right)k \cdot k^{(1)} - 4\, {\rm Tr}\left(b \cdot b^{(1)}\right) k \cdot \epsilon^{(2)} \cdot k \nonumber\\
    &- 8\left(k\cdot \epsilon^{(2)} \cdot b^{(1)}\cdot b\cdot k^{(1)} + k^{(1)}\cdot  b\cdot \epsilon^{(2)} \cdot b^{(1)}\cdot  k - k\cdot \epsilon^{(2)} \cdot b\cdot   b^{(1)} \cdot k \right),
\end{align}
with $k+k^{(1)}=k^{(2)}$ or
\begin{align}
    \left<b^{(1)}\right|V_b\left|h^{(2)}\right> &= 4\, {\rm Tr}\left(\epsilon^{(2)}\cdot b^{(1)} \cdot b\right)k^2 + 8 \,{\rm Tr}\left(\epsilon^{(2)}\cdot b^{(1)} \cdot b\right)k \cdot k^{(2)} - 4\, {\rm Tr}\left(b \cdot b^{(1)}\right) k \cdot \epsilon^{(2)} \cdot k \nonumber\\
    &- 8\left(k\cdot \epsilon^{(2)} \cdot b^{(1)}\cdot b\cdot k^{(2)} + k^{(2)}\cdot  b\cdot \epsilon^{(2)} \cdot b^{(1)}\cdot  k - k\cdot \epsilon^{(2)} \cdot b\cdot   b^{(1)} \cdot k \right),
\end{align}
with $k+k^{(2)}=k^{(1)}$. Note that we have not assumed the mass-shell condition in deriving these three-point functions\footnote{We have, however, assumed transversality.}. These three amplitudes are identical as can be seen using momentum conservation and transversality. Furthermore, for $k^2=0$ (\ref{cfft}) agrees with what is expected for the $3$-function from the string effective action. Finally, the amplitude for three $B$-fields vanishes.
\subsection{Coupling to the dilaton}
In the case of the background metric and the Kalb-Ramond field it is possible to draw from our experience from string theory to make and educated guess on how to couple these fields to the worldline. For the dilaton the situation is different. Indeed, in the textbook formulation of the string worldsheet theory the dilaton couples through the ghost number anomaly to the worldsheet curvature \cite{Banks:1986fu}, for which there is no analogue on the worldline. On the other hand, given that the dilaton is contained in the spectrum, it ought to be able to couple as well. However,  we were not able to formulate a general argument from which this coupling should derive.\footnote{Given that the dilaton sector of 10-dimensional type IIA supergravity can be obtained by Kaluza-Klein reduction of 11-dimensional supergravity (e.g. \cite{Duff:1986hr}) one might expect the same mechanism to work for the coupling at the level of the worldline. However, with the standard Ansatz $\{e^I\}=\{e^a,e^z=e^{\beta\Phi}dz\}$, $a=1,\cdots,d$ and $z$ along the $S^1$, the resulting spin connection has no component along the non-compact dimensions and therefore no coupling for $\Phi$ is induced in this way.} Through a series of trial and error we came up with the following proposal for the supercharges:
\begin{equation}\label{qbarqwithPhi}
q_i:=-i\,e^{\kappa\Phi}\theta^a_i\,e^\mu_a\,\hat D^+_\mu\;,\quad \bar q^i:=-i\,e^{\kappa\Phi}\bar\theta^{a\,i}\,e^\mu_a  {\hat  D}^-_\mu\;,
\end{equation}
with
\begin{equation}
\begin{split}
\hat D^+_\mu:&=\de_\mu+\, \tfrac12\,(\omega_{\mu\,ab}+\kappa\,\Omega_{\mu\,ab})S^{ab}+\de_\mu\Phi\,, \\  \hat D^-_\mu:&=\de_\mu+\tfrac12\,(\omega_{\mu\, ab}+\kappa\,\Omega_{\mu\,ab}) S^{ab}+(d\kappa-1)\,\de_\mu\Phi\,.
\end{split} 
\end{equation}
with $\Omega_{\mu\,ab}=2\,\partial_\nu\Phi\,e^\nu{}_{[a}\, e_{b]\mu}$ and $\kappa\in \mathbb{R}$ parametrizes a Weyl rescaling of the metric. It can be set to zero by a suitable choice of frame which we will later recognize as the "string frame". The geometric interpretation of the deformation $\pm\de_\mu\Phi $ in $\hat D^\pm_\mu$ is less clear but is reminiscent of the Weyl-gauging procedure in \cite{Iorio:1996ad}. Regardless, once $q_i$ is defined via \eqref{qbarqwithPhi}, thus giving the deformation in $\hat D^+_\mu\,$, $\bar q^i\,$, and thus $\hat D_\mu^-\,$, is uniquely fixed by hermiticity: $\bar q^i:=(q_i)^\dagger\,$.

The commutator of these covariant derivatives reads
\begin{equation}
\begin{split}
[\hat{D}^{+}_\mu,\hat{D}^{-}_\nu]&=\tfrac12\,\R_{\mu\nu ab}\, S^{ab}+(d\kappa-2)\,\nabla_\mu\nabla_\nu\Phi+(d\kappa-2)\,\Gamma_{\mu\nu}^\lambda\,\nabla_\lambda\Phi  \;,\\
[\hat{D}^{\pm}_\mu,\hat{D}^{\pm}_\nu]&=\tfrac12\,\R_{\mu\nu ab}\, S^{ab}\;,
\end{split}
\end{equation}
where
\begin{equation}
\R_{\mu\nu\lambda\sigma}=R_{\mu\nu\lambda\sigma}+4\kappa\,\nabla_{[\mu}\nabla_{[\lambda}\Phi\,g_{\sigma]\nu]}+2\kappa^2\,\Big[2\,\de_{[\mu}\Phi\de_{[\lambda}\Phi\,g_{\sigma]\nu]}-(\de\Phi)^2\,g_{\lambda[\mu}\,g_{\nu]\sigma}\Big]\;,    
\end{equation}
and $R_{\mu\nu\lambda\sigma}$ is the usual Riemann tensor. In order to simplify the presentation we set $\kappa=0$ in the following, since we can restore it at any point by a Weyl rescaling of the metric. The algebra of the supercharges then takes the form
\begin{equation}
\begin{split}
\{q_i,q_j\}&=-\tfrac12\,\theta^\mu_i\theta^\nu_j\,R_{\mu\nu ab}\, S^{ab}\;,\\
\{\bar q^i,\bar q^j\}&=-\tfrac12\,\bar\theta^{\mu\,i}\bar\theta^{\nu\,j}\,R_{\mu\nu ab}\, S^{ab}\;,\\
\{q_i,\bar q^j\}&=-\delta_i^j\,\nabla_\Phi^2-\theta^\mu_i\bar\theta^{\nu\,j}\Big[\tfrac12\,R_{\mu\nu ab}\, S^{ab}-2\,\nabla_\mu\nabla_\nu\Phi\Big]\;,
\end{split}    
\end{equation}
Here 
\begin{equation}
\nabla_\Phi^2:=  \frac{1}{\sqrt g}\,\Big(\hat \nabla_\mu-\,\de_\mu\Phi\Big)\,g^{\mu\nu}\sqrt{g}\,\Big(\hat \nabla_\nu+\,\de_\nu\Phi\Big)=\nabla^2+\,\nabla^2\Phi-\,(\de\Phi)^2 
\end{equation}
is the self-adjoint deformed Laplacian. We further define, as before,
\begin{equation}
\triangle:=\nabla_\Phi^2+\Re   
\end{equation}
where $\Re$ is the non-minimal coupling, still to be determined, and the BRST charge takes the usual form
\begin{equation}\label{QwithPhi}
\begin{split}
&Q=c\,\triangle+\bm{D}+\bar\gamma\cdot\gamma\,b\;,\;{\rm with}\\
&\bm{D}:=\bar\gamma\cdot q+\gamma\cdot\bar q=-i\,S^\mu\hat\nabla_\mu-iS_-^\mu\de_\mu\Phi\;,
\end{split}
\end{equation}
with $S_-^\mu:=\bar\gamma\cdot\theta^\mu-\gamma\cdot\bar\theta^\mu$ as defined in the previous subsection.

The first obstruction to nilpotency comes again from
\begin{equation}
\begin{split}
\bm{D}^2+\bar\gamma\cdot\gamma\,\triangle &=-\tfrac14\,S^\mu S^\nu\,R_{\mu\nu\lambda\sigma}\,S^{\lambda\sigma}+2\bar\gamma\cdot\theta^\mu \gamma\cdot\bar\theta^\nu\,\nabla_\mu\nabla_\nu\Phi+\bar\gamma\cdot\gamma\,\Re\\
&\!\!\!\stackrel{{\rm ker}\J_i}{=}\bar\gamma\cdot\theta^\mu \gamma\cdot\bar\theta^\nu \Big[R_{\mu\nu}+2\nabla_\mu\nabla_\nu\Phi\Big]+\bar\gamma\cdot\gamma\,\Re\rvert_{{\rm ker}\J_i}\;,
\end{split}    
\end{equation}
that, in turn, implies the field equation 
\begin{equation}
R_{\mu\nu}+2\,\nabla_\mu\nabla_\nu\Phi=0\;,    
\end{equation}
together with $\bar\gamma\cdot\gamma\,\Re\rvert_{{\rm ker}\J_i}=0\,$. 
The second obstruction is given by 
\begin{equation}
\begin{split}
i[\triangle,\bm{D}]=&\;\;S^\nu\,R_{\mu\nu\lambda\sigma}\,S^{\lambda\sigma}\hat\nabla^\mu+S_-^\nu\,R_{\mu\nu\lambda\sigma}\,S^{\lambda\sigma}\nabla^\mu\Phi +S^\mu(R_{\mu\nu}+2\nabla_\mu\nabla_\nu\Phi)\hat\nabla^\nu\\
&+S_-^\mu(R_{\mu\nu}+2\nabla_\mu\nabla_\nu\Phi)\nabla^\nu\Phi+\nabla_\lambda(R_{\sigma\nu}+2\nabla_\sigma\nabla_\nu\Phi)S^\nu S^{\lambda\sigma}  \\
&-4\nabla_\lambda\nabla_\sigma\nabla_\nu\Phi\,\gamma\cdot\bar\theta^\nu S^{\lambda\sigma}-4\gamma\cdot\bar\theta^\mu\,\nabla_\mu\nabla_\nu\Phi\,\hat\nabla^\nu-2\gamma\cdot\bar\theta^\mu\nabla_\mu(\nabla^2\Phi-(\de\Phi)^2)\\
&+i[\Re,\bm{D}]\;.
\end{split}    
\end{equation}
The only way to make the first terms with the full Riemann tensor vanish is by choosing $\Re=\tfrac14\,R_{\mu\nu\lambda\sigma}\,S^{\mu\nu}S^{\lambda\sigma}+...$ The further requirement $\bar\gamma\cdot\gamma\,\Re\rvert_{{\rm ker}\J_i}=0$ fixes it to
\begin{equation}
\Re=\tfrac14\,R_{\mu\nu\lambda\sigma}\,S^{\mu\nu}S^{\lambda\sigma}-2\nabla_\mu\nabla_\nu\Phi\,\theta^\mu\!\cdot\bar\theta^\nu   \;. 
\end{equation}
Evaluating the obstruction on ${\rm ker}\J_i$ we finally obtain
\begin{align}\label{dilatonsecondobstruction}
 i[\triangle,\bm{D}]\stackrel{{\rm ker}\J_i}{=}&\;\big[ {{\nabla_\lambda }\left( {{R_{\mu\nu }} + 2{\nabla_\mu }{\nabla_\nu }\Phi } \right) - {\nabla_\nu }\left( {{R_{\mu \lambda }} + 2{\nabla_\mu }{\nabla_\lambda }\Phi } \right) + {\nabla_\mu }\left( {{R_{\nu\lambda }} + 2{\nabla_\nu }{\nabla_\lambda }\Phi } \right)} \big] \nonumber\\
 &\times\left( {\theta ^{\nu}\!\cdot \bar \theta^\lambda\, \gamma\cdot\bar\theta^\mu  + \bar\gamma \cdot\theta^\mu\,\theta^\lambda\!\cdot {{\bar \theta }^{\nu}}} \right) \nonumber\\
 =& \;{S^\lambda }{\nabla_\lambda }\left( {{R_{\mu \nu }} + 2{\nabla_\mu }{\nabla_\nu }\Phi } \right)\theta^\mu\!\cdot {{\bar \theta }^{\nu}}  + {S_-^\rho }{\nabla_\mu }\left( {{R_{\nu \rho }} + 2{\nabla_\nu }{\nabla_\rho }\Phi } \right){S^{\mu\nu}} \nonumber\\
 & - {\nabla_\lambda }\left( {R + 2\,\nabla^2\Phi } \right){\gamma }\cdot\bar \theta ^\lambda \,.
\end{align}
This then confirms that the BRST charge is nilpotent when the field equations
\begin{equation}\label{dileq}
R_{\mu\nu}+2\nabla_\mu\nabla_\nu\Phi=0\;,\quad \nabla^2\Phi-2\,\nabla^\mu\Phi\,\nabla_\mu\Phi=0 \; 
\end{equation}
 are satisfied, where the second equation is implied by the first via the Bianchi identity for $R_{\mu\nu}$. These equations are the same one obtains in closed string theory (to lowest order in $\alpha'$) in the string frame. 
 To switch to the Einstein frame one can perform a Weyl transformation on the background metric or, equivalently, introduces a $\kappa$-deformation as above with $\kappa=-\tfrac{2}{d-2}\,$.
 
To be precise, we should point out that the obstructions to $Q^2=0$ only imply the weaker condition
  \begin{equation}
\nabla^2\Phi-2\,(\nabla\Phi)^2=K      \;,
  \end{equation}
 for any real constant $K\,$. Even in string theory, the field equation for the dilaton mostly descends as a Bianchi consistency condition for the other couplings \cite{Curci:1986hi,Callan:1989nz}. The constant $K\,$, in the string framework, is related to the total central charge of the conformal field theory. Demanding zero total central charge one has $K\propto\frac{d-d_{\rm crit}}{\alpha'}\,$, so that $K=0$ for critical strings. The worldline theory, on the other hand, poses no constraints on the value of $K$. Here we choose $K=0$ by demanding that a constant dilaton be a solution in flat spacetime.  
 
 Finally, note that one can further constrain the Hilbert space to allow for a cosmological constant, at the price of projecting out the $B$-field, as explained in \cite{Bonezzi:2018box}. In The BRST operator it just amounts to the constant shift $\Re\to\Re+2\lambda\,$. This produces Einstein gravity with cosmological constant coupled to a scalar field:
 \begin{equation}
    R_{\mu\nu}-\lambda g_{\mu\nu}+2\nabla_\mu\nabla_\nu\Phi=0\;,\quad \nabla^2\Phi-2\,\nabla^\mu\Phi\,\nabla_\mu\Phi + 2\lambda \Phi=0 \;, 
 \end{equation}
 whose field equations can be derived from the spacetime effective action
 \begin{equation}
S=\frac{1}{2\kappa^2}\int d^dx\,\sqrt{-g}\,e^{-2\Phi}\Big[R+4\,g^{\mu\nu}\de_\mu\Phi\,\de_\nu\Phi+4\lambda\,\Phi+(2-d)\lambda\Big] \;.    
 \end{equation}


\subsubsection{Dilaton vertex operator}
The construction of the dilaton state involves an extra complication as compared to the $B$-field or the graviton. By expanding $Q$ around flat space to first order in $\Phi=\sigma\,$, \emph{i.e.}
\begin{equation}
Q=c\,(\Box+W_{\rm I}(\sigma))-iS^\mu\de_\mu+W_{\rm II}(\sigma)+\bar\gamma\cdot\gamma\,b\;,
\end{equation}
the dilaton state is given by 
\begin{equation}\label{QD2}
 \ket{\sigma}=   W_{\rm II}\ket{\xi}_{\eta_{\mu\nu}}-iS^\mu\de_\mu\ket{\xi}_{g_{\mu\nu}}
\end{equation}
with 
\begin{equation}
\ket{\xi}_{g_{\mu\nu}}:=\xi_\mu\,(\theta_1^\mu\beta_2-\theta_2^\mu\beta_1)\ket{0}_{g_{\mu\nu}}\,.    
\end{equation}
Here we used that for a Weyl deformed metric, $g_{\mu\nu}=e^{2\omega}\eta_{\mu\nu}$, the normalized vacuum wave function is given by $\ket{0}_{g_{\mu\nu}}=\frac{1}{|g|^{1/4}}$ with  ${}_{g_{\mu\nu}}\bra{0}\!\!\ket{0}_{g_{\mu\nu}}=1$. 
The second term in \eqref{QD2} then contributes because $\ket{0}_{g_{\mu\nu}}$ is annihilated by $p_\mu$ rather than $\partial_\mu$.\footnote{Note that this extra term is pure gauge but the gauge transformation is non-local. For the transverse graviton vertex this term does not contribute because $\omega=0$ for a linearized, transverse graviton.} We then have, to first order in the Weyl parameter $\omega\,$,  
\begin{eqnarray}
W_{\rm II}\ket{\xi}_{\eta_{\mu\nu}}-iS^\mu\de_\mu\ket{\xi}_{g_{\mu\nu}}&=&-i\left[(\de_\mu\sigma\,\xi_\nu+   \de_\nu\sigma\,\xi_\mu)-\left(1+\tfrac{d}{2}\right)(\de_\mu\omega\,\xi_\nu+   \de_\nu\omega\,\xi_\mu)\right]\theta_1^\mu \theta_2^\nu \ket{0}\nonumber\\
&&+i \left[\de_\mu\sigma\,\xi^\mu+\left(1-\tfrac{d}{2}\right)\de_\mu\omega\,\xi^\mu\right]\left(\gamma_1\beta_2-\gamma_2\beta_1\right)\ket{0}\nonumber\\
&&-2i\,\de_\mu\omega\,\xi^\mu\,\theta_1\!\cdot\theta_2\ket{0}\;.
\end{eqnarray}
For $\omega=0$, we get 
\begin{eqnarray}\label{newstdil}
W_{\rm II}\ket{\xi}&=&-i(\de_\mu\sigma\,\xi_\nu+   \de_\nu\sigma\,\xi_\mu)\theta_1^\mu \theta_2^\nu\ket{0} +i\,\de_\mu\sigma\,\xi^\mu(\gamma_1\beta_2-\gamma_2\beta_1)\ket{0}\,.
\end{eqnarray}
However, this is inconsistent with the on-shell condition (\ref{dileq}) which implies a non-vanishing Ricci tensor. That is, an infintesimal shift in the dilaton background implies a shift in the metric as well through $\omega=\frac{2}{d-2}\sigma$ an thus,
\begin{eqnarray}\label{ENew}
W_{\rm II}\ket{\xi}_{\eta_{\mu\nu}}-iS^\mu\de_\mu\ket{\xi}_{g_{\mu\nu}=(1+2\omega)\eta_{\mu\nu}}&=&\frac{4i}{d-2}\left[(\de_\mu\sigma\,\xi_\nu+ \de_\nu\sigma\,\xi_\mu)\theta_1^\mu \theta_2^\nu-\de_\mu\sigma\,\xi^\mu\;\theta_1\!\cdot\theta_2\right]\ket{0}_{\eta_{\mu\nu}}
\end{eqnarray}
which is the familiar dilaton vertex in string theory. 

Before closing this subsection we would like to mention that there is an alternative representation of the unintegrated dilaton vertex \cite{Kataoka:1990ga} in terms of the superghosts, which survives the reduction to the worldline and which we present in the appendix. 

\subsection{Fully coupled system}

We are now ready to couple the model simultaneously to all backgrounds. The deformed covariant derivatives $\hat D_\mu$ are the same as in the $B$-field section, namely
\begin{equation}
\hat D_\mu=\hat\nabla_\mu+\tfrac12\,H_{\mu ab}\, T^{ab}    
\end{equation}
and the supercharges are given by
\begin{equation}
q_i=-i\,\theta^\mu_i(\hat D_\mu+\de_\mu\Phi)\;,\quad \bar q^i=-i\,\bar \theta^{\mu i}(\hat D_\mu-\de_\mu\Phi)\;.    
\end{equation}
The superalgebra reads
\begin{equation}
\begin{split}
&\{q_i,q_j\}=-\theta^\mu_i\theta^\nu_j\,\bm{C}_{\mu\nu}-(\theta^\mu_i\vartheta^\nu_j+\vartheta^\mu_i\theta^\nu_j)H_{\mu\nu}{}^\lambda(\hat D_\lambda+\de_\lambda\Phi)\;,\\[2mm]
&\{\bar q^i,\bar q^j\}=-\bar\theta^{\mu\,i}\bar\theta^{\nu\,j}\,\bm{C}_{\mu\nu}-(\bar\theta^{\mu\,i}\bar\vartheta^{\nu\,j}+\bar\vartheta^{\mu\,i}\bar\theta^{\nu\,j})H_{\mu\nu}{}^\lambda(\hat D_\lambda-\de_\lambda\Phi)\;,\\[2mm]
&\{q_i,\bar q^j\}=-\delta_i^j\,D_\Phi^2-\theta^\mu_i\bar\theta^{\nu\,j}\,[\bm{C}_{\mu\nu}-2\nabla_\mu\nabla_\nu\Phi]-(\theta^\mu_i\bar\vartheta^{\nu\,j}+\vartheta^\mu_i\bar\theta^{\nu\,j})H_{\mu\nu}{}^\lambda\hat D_\lambda+(\theta^\mu_i\bar\vartheta^{\nu\,j}-\vartheta^\mu_i\bar\theta^{\nu\,j})H_{\mu\nu}{}^\lambda\,\de_\lambda\Phi\;,
\end{split}    
\end{equation}
where now
\begin{equation}
D_\Phi^2:=D^2+\nabla^2\Phi-(\de\Phi)^2\;,\quad D^2=g^{\mu\nu}(\hat D_\mu\hat D_\nu-\Gamma_{\mu\nu}^\lambda\,\hat D_\lambda)    
\end{equation}
and we recall the notation
\begin{equation}
\begin{split}
\bm{C}_{\mu\nu}&:=[\hat D_\mu, \hat D_\nu]=\tfrac12\,\R_{\mu\nu ab}\, S^{ab}+\nabla_{[\mu}H_{\nu]ab}\,T^{ab}  \;,\\[2mm]
\R_{\mu\nu\lambda\sigma}&:=R_{\mu\nu\lambda\sigma}-H_{\mu\lambda}{}^\alpha H_{\nu\sigma\alpha}+H_{\nu\lambda}{}^\alpha H_{\mu\sigma\alpha}\;.
\end{split}
\end{equation}
We also rewrite for convenience the definition of all the relevant ghost-valued vectors:
\begin{equation}
\begin{split}
S^\mu&:=\bar\gamma\cdot\theta^\mu+\gamma\cdot\bar\theta^\mu\;,\quad T^\mu:=\bar\gamma\cdot\vartheta^\mu+\gamma\cdot\bar\vartheta^\mu\;,\\
S_-^\mu&:=\bar\gamma\cdot\theta^\mu-\gamma\cdot\bar\theta^\mu \;,\quad T_-^\mu:=\bar\gamma\cdot\vartheta^\mu-\gamma\cdot\bar\vartheta^\mu\;.  
\end{split}    
\end{equation}
For the BRST operator we make the Ansatz
\begin{equation}\label{QwithBPhi}
\begin{split}
&Q=c\,\triangle+\bm{D}+\bar\gamma\cdot\gamma\,b\;,\;{\rm with}\\
&\bm{D}:=\bar\gamma\cdot q+\gamma\cdot\bar q=-i\,S^\mu\hat D_\mu-i\,S_-^\mu\,\de_\mu\Phi\quad{\rm and}\quad \triangle:=D_\Phi^2+\Re\;.
\end{split}
\end{equation}
The first obstruction to the nilpotency of $Q$ is given by
\begin{equation}\label{first obstruction with BPhi}
\begin{split}
\bm{D}^2+\bar\gamma\cdot\gamma\,\triangle&=-\tfrac12\,S^\mu S^\nu\,\bm{C}_{\mu\nu}-S^\mu T^\nu\,H_{\mu\nu}{}^\lambda\hat D_\lambda-S^\mu T_-^\nu\,H_{\mu\nu}{}^\lambda\,\de_\lambda\Phi +2\bar\gamma\cdot\theta^\mu\gamma\cdot\bar\theta^\nu\,\nabla_\mu\nabla_\nu\Phi+\bar\gamma\cdot\gamma\,\Re \\[2mm]
&\!\!\!\stackrel{\ker\J_i}{=} \bar\gamma\cdot\theta^\mu\gamma\cdot\bar\theta^\nu\,\big[\R_{\mu\nu}+2\,\nabla_\mu\nabla_\nu\Phi\big]-\bar\gamma\cdot\theta^\mu\gamma\cdot\bar\vartheta^\nu\,\big[\nabla^\lambda H_{\lambda\mu\nu}-2\,H_{\mu\nu\lambda}\,\nabla^\lambda\Phi\big]+\bar\gamma\cdot\gamma\,\Re\rvert_{\ker \J_i}\;.
\end{split}    
\end{equation}
The requirement for this obstruction to vanish leads to the following background field equations
\begin{equation}\label{Complete_Field_Equations}
R_{\mu\nu}+2\,\nabla_\mu\nabla_\nu\Phi-H_{\mu\lambda\sigma}\,H_\nu{}^{\lambda\sigma}=0\;,\quad \nabla^\lambda H_{\lambda\mu\nu}-2\,H_{\mu\nu\lambda}\,\nabla^\lambda\Phi=0\,.    
\end{equation}
From consistency of the above equations with the Bianchi identities one finds a third equation,
\begin{equation}\label{dilatonequation}
 {\nabla^\mu }{\nabla_\mu }\Phi  - 2\,{\nabla^\mu }\Phi {\nabla_\mu }\Phi+\tfrac{1}{3}{H^{\mu \nu \sigma }}{H_{\mu \nu \sigma }} = 0\;, 
\end{equation}
where an arbitrary constant on the right hand side has been set to zero according to the discussion in the previous subsection, \emph{i.e.} by demanding that a constant dilaton be a solution for flat space with $H_{\mu\nu\lambda}=0\,$. 
After rescaling, $H_{\mu\nu\lambda}\rightarrow \frac{1}{2}H_{\mu\nu\lambda}$, these completely reproduce the (lowest order in $\alpha'$) closed string field equations for the massless modes. This is the key result of this paper, showing that quantum consistency of the spinning worldline already produces the effective action of the massless fields in the NS-sector of string theory.

Finally, to show consistency, we need to find the correct non-minimal coupling. As before, we need to impose in addition $\bar\gamma\cdot\gamma\,\Re\rvert_{\ker \J_i}=0\,$ which helps to fix the form of $\Re\,$. To continue we make the following Ansatz for $\triangle$:
\begin{align}
\triangle 
&=D^2_\Phi + \tfrac{1}{2}\,S^{\mu\nu}\bm{C}_{\mu\nu}- \tfrac{1}{2}\,\nabla^\mu H_{\mu\nu\lambda}\,T^{\nu\lambda} - 2\,\nabla_\mu\nabla_\nu \Phi\, \theta^\mu\! \cdot \bar{\theta}^\nu,
\end{align}
that coincides with the sum of the various contributions to $\Re$ found in the previous sections for the separate backgrounds.

It now remains to show that $[\triangle, \bm{D}] = \bar{\gamma}^i[\triangle,  q_i] + \gamma_i [\triangle,  \bar{q}^i]$ vanishes provided the background field equations \eqref{Complete_Field_Equations} and \eqref{dilatonequation} are satisfied. In order to keep the result readable, we denote the field equations for the metric and $B$-field as
\begin{equation}
\G_{\mu\nu}:=R_{\mu\nu}+2\,\nabla_\mu\nabla_\nu\Phi-H_{\mu\lambda\sigma}H_\nu{}^{\lambda\sigma}\;,\quad \E_{\mu\nu}:=\nabla^\lambda H_{\lambda\mu\nu}-2\,H_{\mu\nu\lambda}\,\nabla^\lambda\Phi\;.    
\end{equation}
The final obstruction, evaluated on ${\rm ker}\J_i\,$, reads
\begin{equation}
\begin{split}
i[ \triangle,\bm{D} ] 
  & \stackrel{{\rm ker}\J_i}{=}  S^\rho_-\,\Big[(\nabla_\mu \G_{\nu\rho }+ {H_{\rho\lambda \mu}}\,\E^\lambda{}_\nu ) {S^{\mu\nu}}+( {\nabla_\mu}\E_{\nu\rho}-\G_{\mu\lambda}\,H^\lambda{}_{\rho\nu})\,{T^{\mu\nu}}\Big]\\[2mm]
  &\hspace{8mm}+S^\rho\,\big({\nabla_\rho }\G_{\mu\nu}\,\theta^\mu\!\cdot\bar\theta^\nu+2\,\G_{\mu\lambda}\,H^\lambda{}_{\rho\nu}\,{\theta ^{(\mu}}\! \cdot{{\bar \vartheta }^{\nu)}}\big)+\E_{\mu\nu}\big(T^\mu_-\hat D^\nu+T^\mu\nabla^\nu\Phi\big)\\[2mm]
  &\hspace{8mm}+2\,\gamma\cdot\bar\theta^\mu\,\Big[ 2\,\G_{\mu\nu}\,\nabla^\nu\Phi
  +{\nabla_\mu}\left(\nabla^2\Phi-2\,(\nabla\Phi)^2+\tfrac13\,H^2  \right)
  - H_\mu{}^{\nu\lambda}\,\E_{\nu\lambda} - {\nabla^\nu }\G_{\mu\nu}\Big]\;,
\end{split}    
\end{equation}
that clearly vanishes upon putting the background on-shell, without any further constraints. The above expression also makes clear that the dilaton equation appears only differentiated, as a consequence of Bianchi identities.

\section{Conclusions}\label{con}
In this paper we have shown that the low energy effective action of all massless fields of NS-NS sector of type II string theory is already implied by the quantum consistency of the supersymmetric spinning particle. 
Given that the massless NS-spectrum is reproduced by the spinning particle, it is expected that background fields of the same type should be able to couple to the worldline. In addition, the $\mathcal{N}=4$ worldline multiplet can be shown \cite{Bonezzi:2018box} to be related to the center of mass of the string together with the oscillators of lowest frequencies of its fermionic superpartners.

It was shown in \cite{Dai:2008bh} for the case of the $\N=2$ spinning particle coupled to Yang-Mills, and in \cite{Bonezzi:2018box} for $\N=4$ coupled to pure gravity
 that quantum consistency of first-quantized systems coupled to non-trivial background fields\footnote{It should be specified that this seems to be the case only when the background fields are the ones corresponding to the quantum states of the system.} ``predicts'' the spacetime dynamics of the latter. Our result then confirms that the spinning particle already determines completely the spacetime low energy effective action of the string.
This feature is thus not exclusive to string theory (whose exclusive property is rather to provide a UV completion), but a rather general property of first-quantized models whose BRST operator encodes spacetime gauge symmetries \cite{Grigoriev:2019ojp}. It would be interesting to clarify this point in full generality. Along these lines, for instance,  it should be possible to derive Einstein's equations also by considering the $\N=3$ particle \cite{Corradini:2010ia} (describing a spin $\frac{3}{2}$ gravitino) coupled to a curved background.

  We do not see any obvious obstruction to extend the present treatment to include higher modes of the string. It would be interesting to determine their effect on the constraint algebra and spacetime effective action. On the other hand, conformal invariance plays no role in the present analysis which means, in particular, that the dimension of spacetime is not determined. 

Another important feature of our construction of the BRST charge is not assuming any particular background. As such, this construction is truly background independent, although with a caveat: the $\N=4$ spinning particle, as a perturbative quantum theory, is consistent only for on-shell background fields. For instance, this has been tested in \cite{Bastianelli:2019xhi}, where the one-loop divergencies of pure quantum gravity could be reproduced, by using the maximally projected $\N=4$ particle of \cite{Bonezzi:2018box}, only for on-shell Einstein metrics. 

There are a number of possible extensions of the construction presented here, such as including the Ramond sector, which corresponds to space-time fermions\footnote{See for instance \cite{Ahmadiniaz:2020wlm} for a recent attempt to an efficient worldline description of external fermion lines.}, as well as considering higher $\mathcal{N}>4$ which correspond to higher spin particles \cite{Henneaux:1987cp,Howe:1988ft,Bekaert:2002dt,Bastianelli:2007pv,Bastianelli:2008nm,Bastianelli:2012bn} and possibly pure spinors \cite{Berkovits:2002zk}. We hope to return to some of these extensions in future work. Another interesting question is to develop a better understanding of the coupling of the worldline to the dilaton which appears presently in a somewhat ad-hoc manner by trial and error. This might give further insight on the topological properties of worldline graphs\footnote{We would like to thank Warren Siegel for helpful comments on this issue.}. 

\appendix

\section{Alternative dilaton coupling}
Starting form the "string field" (\ref{BV superfield}) and setting the linearized graviton to zero in the Einstein frame we get $\varphi_{\mu\nu}=\frac{1}{d-2}\,\sigma\,\eta_{\mu\nu}$. Then (with $A^\pm_\mu=0$) the dilaton state takes the form 
\begin{equation}\label{sft12}
\ket{\psi}=\tfrac{1}{d-2}\,\sigma\,(\,\theta_1\!\cdot\theta_2+\gamma_1\beta_2-\gamma_2\beta_1)\ket{0}\;,
\end{equation}
which is in agreement with the dilaton vertex for the type II string proposed in \cite{Kataoka:1990ga}. In order to reproduce this state as the linear variation of the BRST charge $Q$ acting on the diffeomorphism ghost state we may take 
\begin{align}
    Q&=c\,\triangle + \bar{\gamma}^iq_i + \gamma_i\bar{q}^{i} +\gamma_i\bar{\gamma}^ib \nonumber\\
    &+ 2c\left(\mathcal{G}\,\bar{\theta}^{1\mu}\bar{\theta}^{2\nu}+\theta_1^\mu \theta_2^\nu \mathcal\,{T}r\right)\nabla_\mu\nabla_\nu\Phi -2i\left(\mathcal{G}\,\bar{\gamma}^{[1}\bar{\theta}^{2]\mu} + \gamma_{[1}\theta_{2]}^\mu \mathcal{T}r\right)\,\nabla_\mu\Phi\;,
\end{align}
where $\triangle$ and the supercharges are the ones given in \eqref{curvedQ} for pure gravity, and
\begin{equation}
\begin{split}
&{\mathcal{T}r} \equiv -\tfrac{1}{2}\left( {i{\J_{12}} - i{\J_{34}} - {\J_{23}} + {\J_{14}}} \right) = {{\bar \theta }^1}\!\cdot{{\bar \theta }^2} - {{\bar \beta }^1}{{\bar \gamma }^2} + {{\bar \beta }^2}{{\bar \gamma }^1},\\
&\mathcal{G} \equiv -\tfrac{1}{2}\left( {i{\J_{12}} - i{\J_{34}} + {\J_{23}} - {\J_{14}}} \right) = {\theta _1}\!\cdot{\theta _2} - {\beta _1}{\gamma _2} + {\beta _2}{\gamma _1}.
\end{split}
\end{equation}
are $SO(4)$ generators (see  \cite{Bonezzi:2018box} for more details). The novel feature here is that as opposed to the standard BRST procedure both the Hamiltonian and the supercharges have a manifest dependence on the superghosts and anti-ghosts. It then appears that the corresponding BRST charge in a dilaton background does not derive form Dirac constraints in the standard manner.\footnote{We can, however, not exclude the existence of a similarity transformation that  takes the BRST charge in the standard form.} 

Squaring the BRST operator we then find,
\begin{align} \label{Qdilaton2}
{Q^2} &=  - ic\left( {{{\bar \gamma }}\cdot\theta ^\lambda\theta^\mu\!\cdot{{\bar \theta }^{\nu}} + \theta^\nu\!\cdot{{\bar \theta }^{\mu}}{\gamma }\cdot\bar \theta ^\lambda} \right)\nonumber\\
&\times\Big[ 
{\nabla_\mu}\big( R_{\nu\lambda} - \left( d - 2 \right)\de_\nu\Phi\de_\lambda\Phi \big)
 - {\nabla_\nu}\big( R_{\mu\lambda} - \left( d - 2 \right)\de_\mu\Phi \de_\lambda\Phi \big)
 + {\nabla_\lambda}\big( {R_{\mu\nu}} - \left( {d - 2} \right)\de_\mu\Phi\de_\nu\Phi \big)
 \Big] \nonumber\\
 &- {\gamma}\cdot{{\bar \gamma }}\big( {R_{\mu\nu}} - \left( {d - 2} \right)\de_\mu\Phi\de_\nu\Phi \big)\theta ^\mu\!\cdot{{\bar \theta }^{\nu}} +\bar\gamma\cdot\theta^\mu\,\gamma\cdot\bar\theta^\nu\big(R_{\mu\nu}-(d-2)\de_\mu\Phi\de_\nu\Phi\big) \nonumber\\
& + 4ic\,\Big[\G\,\bar\theta^{\mu[1}\bar\gamma^{2]}-\theta^\mu_{[1}\gamma_{2]}\,{\cal T}r\Big]\left(R_{\mu\nu}\,\nabla^\nu\Phi+\tfrac12\,\nabla_\mu\nabla^2\Phi-\nabla^2\Phi\,\nabla_\mu\Phi\right).
\end{align}
If we impose
\begin{align} \label{EOM-g+d}
    &R_{\mu\nu}   - (d-2)\partial_\mu \Phi \partial_\nu \Phi=0,
\end{align}
then all terms except the last one in \eqref{Qdilaton2} vanish. Combining eq (\ref{EOM-g+d}) with the Bianchi identity, $\nabla^\mu R_{\mu\nu}=\frac12 \nabla_\nu R$, gives,
\begin{align}\label{EOM-d}
    \nabla^2 \Phi=0\;.
\end{align}
Thus, in order for $Q^2$ to vanish we must have $R_{\mu\nu}\nabla^\nu\Phi=0\,$. Using equation  (\ref{EOM-g+d}) and its trace one further obtains
\begin{align}
  0=  R_{\mu\nu}\,\nabla^\nu\Phi = R\,\nabla_\mu\Phi\;,
\end{align}
which, in turn, implies that on top of (\ref{EOM-g+d}) we need to impose that the Ricci scalar has to vanish:
\begin{align}\label{constraint}
    R=0\;.
\end{align}
Clearly, these conditions are stronger that what is implied by (\ref{dileq}). Still, there exist non-trivial solutions to this set of equations. Indeed, one can check that the following solution is compatible  with the above equations,
\begin{align}
     &ds^2 = dudv-H_{ab}\left(u\right)x^ax^bdu^2+dx^adx_a \\
     &\Phi=\Phi\left(u\right),\nonumber
\end{align}
with  $H^a_a=R_{uu}=(d-2)\partial_u\Phi\partial_u\Phi$. Moreover, one can check that $R=0$.
The above metric  solution characterizes non-linear plane waves (e.g. \cite{Stephani:2003tm}).

\section*{Acknowledgments}
We would like to thank Maxim Grigoriev and Warren Siegel for helpful discussions.  The work of I.S. was supported, in parts, by the DFG Excellence cluster ORIGINS. The work of R.B. is supported by the ERC Consolidator Grant ``Symmetries \& Cosmology''. 

\bibliographystyle{unsrt}
\bibliography{ref}
\end{document}